\documentclass[12pt,a4paper,notitlepage]{article}

\usepackage{amsfonts}
\usepackage{amsmath}
\usepackage{amssymb}
\usepackage{amsthm}
\usepackage{color}
\usepackage[utf8]{inputenc}
\usepackage{graphicx}
\usepackage{hyperref}
\hypersetup{
  colorlinks,
  citecolor=blue,
  linkcolor=blue,
  urlcolor=blue}
\usepackage{mathrsfs}
\usepackage{multirow}
\usepackage{xurl}
\usepackage[square, numbers, comma, sort&compress]{natbib}
\usepackage[format=hang]{subcaption}
\usepackage{times}
\usepackage{upgreek}
\usepackage{bm}
\usepackage{import}
\usepackage{bbm}
\usepackage{mathtools}
\usepackage[nameinlink]{cleveref}
\usepackage{multirow}
\usepackage{float}
\usepackage{setspace}
\usepackage{array}
\usepackage{newdefs}
\AtBeginDocument{} 
\usepackage{blindtext}
\usepackage{pgf}
\usepackage{tikz}
\usetikzlibrary{shapes.geometric, arrows}
\usepackage{pgfplots}
\pgfplotsset{compat=newest}
\usepackage{pgfplotstable}
\usepgfplotslibrary{groupplots}
\usepackage{color}
\usetikzlibrary{calc,backgrounds} 
\usepackage[symbol]{footmisc}
\usepackage{algorithm}
\usepackage{algpseudocode}
\algnewcommand{\LineComment}[1]{\State \(\triangleright\) #1}

\DeclareMathOperator{\trace}{tr}

\DeclareMathOperator{\deviatoric}{dev}
\newcommand{\dev}[1]{{\deviatoric\left(#1 \right)}}
\newcommand{\nel}{n_{\mathrm{el}}}
\newcommand{\spacee}{{\hspace{0.055em}}}
\newcommand*\circled[1]{\tikz[baseline=(char.base)]{
            \node[shape=circle,draw,inner sep=2pt] (char) {#1};}}

\definecolor{rwthblue}{RGB}{0,84,159}      
\definecolor{rwthlightblue}{RGB}{142,186,229}   
\definecolor{rwthpetrol}{RGB}{0,97,101}      
\definecolor{rwth4}{RGB}{0,152,161}     
\definecolor{rwthgreen}{RGB}{87,171,39}     
\definecolor{rwth6}{RGB}{189,205,0}     
\definecolor{rwth7}{RGB}{255,237,0}     
\definecolor{rwthorange}{RGB}{246,168,0}     
\definecolor{rwth9}{RGB}{227,0,102}     
\definecolor{rwthred}{RGB}{204,7,30}     
\definecolor{rwth11}{RGB}{161,16,53}    
\definecolor{rwth12}{RGB}{97,33,88}     
\definecolor{rwth13}{RGB}{122,111,172}  

\date{}

\graphicspath{{./images/}}

\begin{document}
\pgfkeys{/pgf/number format/.cd,1000 sep={}}

\author{\large{$\text{Njomza Pacolli}^{\mathrm{a},*}$, $\text{Ahmad Awad}^{\mathrm{a}}$, $\text{Jannick Kehls}^{\mathrm{a}}$, $\text{Bjorn Sauren}^{\mathrm{b}}$ , } \\$\text{Sven Klinkel}^{\mathrm{b}}$ , {$\text{Stefanie Reese}^{\mathrm{a,c}}$ , $\text{Hagen Holthusen}^{\mathrm{a}}$ }\\[0.5cm]
  \hspace*{-0.1cm}
  \normalsize{\em ${{}^\mathrm{a}}$RWTH Aachen
  University, Institute of Applied Mechanics}\\
  \normalsize{\em ${{}^\mathrm{b}}$RWTH Aachen
  University, Chair of Structural Analysis and Dynamics}\\
  \normalsize{\em Mies-van-der-Rohe-Str.\ 1, 52074 Aachen, Germany}\\
  \normalsize{\em ${{}^\mathrm{c}}$University of Siegen, Adolf-Reichwein-Straße 2a, 57076 Siegen, Germany}\\
  [0.25cm]\\
}
\title{\LARGE An enhanced single Gaussian point continuum finite element formulation using automatic differentiation}
\maketitle

\small
{\bf Abstract.}
This contribution presents an improved low-order 3D finite element formulation with hourglass stabilization using automatic differentiation (AD). 
Here, the former Q1STc formulation is enhanced by an approximation-free computation of the inverse Jacobian. To this end, AD tools automate the computation and allow a direct evaluation of the inverse Jacobian, bypassing the need for a Taylor series expansion. Thus, the enhanced version, Q1STc+, is introduced. Numerical examples are conducted to compare the performance of both element formulations for finite strain applications, with particular focus on distorted meshes. Moreover, the performance of the new element formulation for an elasto-plastic material is investigated. To validate the obtained results, a volumetric locking-free element based on scaled boundary parametrization is used. Both the implementation of the element routine Q1STc+ and the corresponding material subroutine are made accessible to the public at \url{https://doi.org/10.5281/zenodo.14259791}

\footnotetext[1]{Corresponding author\\ njomza.pacolli@ifam.rwth-aachen.de}

\vspace*{0.3cm}
{\bf Keywords:} {hourglass stabilization, algorithmic differentiation, plasticity, reduced integration, locking-free formulation, finite element technology, automatic differentiation}

\normalsize

\section{Introduction}

The use and application of low-order 3D finite element formulations has proven to be robust and computationally efficient. However, the use of these methods may lead to shear locking in bending-dominated problems and to volumetric locking for nearly incompressible materials in mechanical systems, which causes the system to behave too stiffly. To overcome these deficits, concepts such as the enhanced assumed strain method (EAS) have been introduced, which use a two-field functional based on the Hu-Washizu variational principle. Such element formulations have proven to deliver plausible results for bending dominated problems and for materials that reach the
incompressible limit. 
\subsection{State of the art}
\textbf{Reduced integraton with hourglass stabilization.} The EAS method was first developed and further improved by \citet{simo_geometrically_1992} and \citet{ simo_improved_1993}. Several authors extended and further developed the original idea, e.g.~\cite{klinkel_geometrical_1997, cao_3d_2002, areias_analysis_2003}, for various applications such as (solid) shells, e.g.~\cite{betsch_4-node_1996, betsch_numerical_1999, bischoff_shear_1997, buchter_three-dimensional_1994, cardoso_enhanced_2008, klinkel_robust_2006}. Further approaches and investigations in combination with the EAS concept have also been employed in works of \citet{korelc_efficient_1996}, \citet{sousa_new_2003}, \citet{valente_enhanced_2004}, \citet{bieber_artificial_2023}, \citet{pfefferkorn_hourglassing-_2023} and \citet{pfefferkorn_improving_2021}. Besides this,\citet{wriggers_note_1996} showed that nonlinear enhanced strain elements subjected to a homogeneous compression state denote a rank deficiency in the tangent matrix. Similar observations were also made in~\cite{de_souza_neto_remarks_1995} and~\cite{crisfield_enhanced_1995}. A possible solution would be to use a higher order of integration. However, this does not work for a homogeneous compression state, since the stresses, strains and the Jacobian remain constant. In this case, a low order integration would already suffice to ensure full integration. Works of \citet{korelc_consistent_1996} as well as \citet{glaser_formulation_1997} improved the formulation by interpolating the enhanced strain in a way that the hourglass instability occuring for uniaxial compression is eliminated. \par 
\citet{reese_new_1999} developed a new stabilization technique in nonlinear elasticity and showed that it is sufficient to base the computation of the stabilization matrix only on the material part of the tangent operator. In addition to that, the concept of the equivalent parallelogram derived by \citet{arunakirinathar_geometrical_1995, arunakirinathar_further_1995} was used in order to be able to determine the stabilization factors. Based on~\cite{reese_new_1999}, \citet{reese_stabilization_2000} derived a formulation which can be applied for general deformation states and all arbitrary element distortions. Here, hourglass instabilities can be detected on element level. The element formulation is based on the idea of employing the split of the tangent matrix into a constant and hourglass part. Because of this, an eigenvalue analysis is carried out and in case of instabilities, the stabilization matrix is updated. This approach combined the enhanced strain concept with the stabilization technique derived in \citet{belytschko1984hourglass}, where the so-called stabilization vector $\bm{\gamma}$ was introduced. This formulation was further improved by \citet{reese_consistent_2003, reese_physically_2005}.\par
The use of a one-point quadrature in comparison to full quadrature lies in efficiency in a sense that the computation time is reduced and, depending on the constitutive model, history variables need to be stored for only one integration point. However, hourglass control is necessary in order to obtain plausible results. Reduced integration without hourglass stabilization can cause rank deficiency in the stiffness matrix and results in an element response that is too soft. Pioneering works of e.g. \citet{hughes_equivalence_1977} and \citet{belytschko1984hourglass} contributed towards further development of reduced integration techniques and hourglass stabilization methods based on mixed variational principles. \citet{bonet_uniform_1995} extended hourglass stabilization techniques by introducing an artificial hourglass control. \citet{schulz_finite_1985} carried out a Taylor series expansion of the stress with respect to the center of the element, ensuring hourglass control due to the retention of the first and second term of the expansion. This idea was then employed by works of e.g. \citet{reese_consistent_2003, reese_physically_2005, reese_large_2007} as well as \citet{juhre_reduced_2010}. This allows for the hourglass parts to be computed directly instead of solving the system by linearizing the variational principle as shown in~\cite{reese_stabilization_2000}. Moreover, a simpler and more efficient element formulation is obtained. Works of \citet{schwarze_reduced_2009,schwarze_reduced_2011}, \citet{frischkorn_solid-beam_2013} and \citet{barfusz_single_2021, barfusz_reduced_2021} extended this idea by also taking a Taylor series expansion of the inverse Jacobian to take the geometry into account in a more realistic manner. Other possible solutions to overcome the problem of locking and forms of hourglass stabilization were discussed in e.g.~\cite{liu_multiple-quadrature_1998, puso_highly_2000, bieber_variational_2018} and~\cite{mohseni-mofidi_application_2021, hao_stabilized_2023}, respectively. \par
\textbf{Recent developments.}
The virtual element method (VEM) has been established recently and was first introduced by \citet{veiga_basic_2012}. Similar to the concept of reduced integration, virtual elements must be stabilized as well to avoid a rank deficiency in the element formulation. \citet{wriggers_efficient_2017} introduced the idea of approximating the displacement field by an interior triangular finite element mesh to stabilize the formulation. \citet{cangiani_hourglass_2015} connected the VEM to classical hourglass stabilization techniques. The advantage over using the classical finite element method lies in the flexibility when it comes to mesh generation and regeneration, since non-convex elements and elements with an arbitrary number of nodes are possible. Examples of other works regarding this field are~\cite{da2013virtual, gain2014virtual, wriggers2020virtual, cihan_virtual_2022, bohm_mixed_2023}. The capability to handle polygonal and polyhedral elements is not merely a benefit of the VEM. Recent developments in the scaled boundary finite element method \cite{song_scaled_2018}, such as formulations proposed by \citet{klinkel_finite_2019}, \citet{sauren_mixed_2023} or \citet{sauren_mesh_2024}, are able to discretize polygonal finite elements at large strains and do not require additional stabilization.  \par
\textbf{Automatic differentiation.} 
In computational mechanics, the development of a robust and efficient numerical implementation plays an important role. Furthermore, since usually a large number of derivations are involved, hand calculations can be error-prone and time-consuming. A remedy for this is the introduction of automatic differentiation (AD)~\cite{wengert1964simple, bartholomew-biggs_automatic_nodate}. AD has become essential in fields like optimization, structural mechanics, and material modeling. Unlike numerical differentiation, which approximates derivatives using finite differences, AD provides exact derivatives by propagating derivatives through the computational graph of a program, see e.g.~\cite{griewank2008evaluating}. This makes AD more suitable for problems where high accuracy is required, and manual differentiation is impractical. AD can be implemented in two primary modes: forward mode and backward mode, see e.g.~\cite{korelc2016automation} or~\cite{griewank2008evaluating} for a more detailed explanation. AD tools were made compatible for a number of programming languages such as \textit{FORTRAN}~\cite{hascoet_tapenade_2013} or \textit{Mathematica}~\cite{korelc_multi-language_nodate}. \par 
The present contribution focuses on the implementation of an enhanced element formulation based on a single Gaussian point with hourglass stabilization with help of AD. In this work, the AD tool \textit{AceGen}~\cite{korelc_multi-language_nodate, korelc2016automation} is used for implementation. Other AD tools such as \textit{Tapenade}~\cite{hascoet_tapenade_2013} or \textit{JAX}~\cite{bradbury2018jax}, also present possible alternatives. In \textit{AceGen}, the approach of "Simultaneous Stochastic Simplification of numerical code", see~\cite{korelc_automatic_1997}, is used, where both symbolic and algebraic capabilities of the programming language \textit{Mathematica} are combined. A numerically efficient code is obtained due to \textit{AceGen}'s code optimization techniques and the combination of both the forward and backward mode with a \textit{code-to-code} transformation strategy, e.g.~\cite{griewank2008evaluating}, where a source code suitable for the progamming languages \textit{FORTRAN}, \textit{C/C++} and \textit{Mathematica} is generated~\cite{hudobivnik_closed-form_2016, korelc2016automation}. In comparison, \textit{JAX} dynamically compiles functions at runtime using just-in-time (JIT) compilation for each execution, while \textit{AceGen} (and \textit{Tapenade}) produces precompiled, reusable code, making it especially suitable for large-scale simulations and environments requiring high performance.
\subsection{Hypothesis}
Without the use of AD, it is hardly possible to compute some specific needed quantities, especially when highly nonlinear partial derivatives are involved, as would be the case for the inverse Jacobian. Previous publications of \citet{barfusz_single_2021} or \citet{schwarze_reduced_2011, schwarze_reduced_2009} proposed to approximate the highly nonlinear inverse Jacobian by means of a Taylor series expansion instead of only evaluating it at the element center as was done in \citet{reese_large_2007} or \citet{legay_elastoplastic_2003}. Moreover, \citet{schwarze_reduced_2009} stated that a linear approximation of the inverse Jacobian would be sufficient for the transformation of the contravariant basis to the cartesian one. With the use of AD, this approximation would not be necessary anymore. Consequently, the enhanced formulation is approximation-free regarding the inverse Jacobian. We hypothesize that this improvement will have an influence on the results obtained from numerical examples. The main objective of this contribution is to compare results obtained from the Q1STc formulation derived by \citet{barfusz_single_2021, barfusz_reduced_2021} -- where the geometry is approximated by means of a Taylor series expansion -- with the modified element formulation Q1STc+ without an approximation of the geometry. To do so, certain patch tests, such as the membrane patch test and the patch test for solids, are considered.
\subsection{Outline}
The present contribution is structured as follows. Section~\ref{plasti_theory} presents the elasto-plastic constitutive framework, see~\cite{vladimirov_anisotropic_2008}. Here, the volumetric-isochoric split of the Helmholtz free energy is employed in order to compare results with the locking-free mixed scaled boundary finite element formulation provided by \citet{sauren_mixed_2023, sauren_stability_2024,sauren_mesh_2024}. In Section~\ref{Two-field variational functional}, the classical two-field variational functional used by \citet{simo_geometrically_1992} is recapitulated. Section~\ref{Kinematics} derives the kinematics of the system and shows the interpolation of the needed quantities. Section~\ref{Gauss_point_concept} deals with the concept of reduced integration and the connected Taylor series expansion of the quantities in order to obtain a stable and locking-free element formulation. Section~\ref{Approx_Jaci} explains the main differences between Q1STc and Q1STc+. The discretization of the weak form and the global assembly are shown in Section~\ref{Discretization of the weak form}. It is crucial to point out that throughout the contribution, the use of AD plays an important role, particularly when it comes to obtaining nonlinear derivations such as the inverse Jacobian. Numerical examples are conducted to compare and analyze the performances of Q1STc and Q1STc+, while the results determined with the displacement-pressure element of \cite{sauren_mesh_2024} are used to validate the performance of the modified element formulation Q1STc+. The contribution closes with a conclusion and outlook (Section~\ref{conclusion}). 
\section{Constitutive framework}
\label{plasti_theory}
The following section briefly describes the constitutive framework presented in~\cite{vladimirov_anisotropic_2008} in the context of the co-rotated intermediate configuration, cf.~\cite{holthusen_inelastic_2023}. Here, an elasto-plastic material is considered, specifically constrained to volume-preserving plasticity. Non-volume-preserving plasticity models, e.g. Drucker-Prager plasticity~\cite{drucker_soil_1952}, are out of the scope of this contribution. The main idea of the co-rotated intermediate configuration lies in pulling back all quantities that are normally affected by the rotational non-uniqueness to this very configuration instead of doing a pull-back to the reference configuration, e.g.~\cite{dettmer_theoretical_2004}. Due to this, all needed quantities can be directly computed. Especially when using AD, this proves to be an advantage since these quantities, e.g. stresses, driving forces, can be derived directly from the defined energy.\par
\textbf{Kinematics.}
To account for nonlinear plastic material behavior, the deformation gradient is assumed to be multiplicatively split into an elastic part $\bm{F}_e$ and a plastic part $\bm{F}_p$. In order to model kinematic hardening properly, an additional split of the plastic part of the deformation gradient is carried out~\cite{lion_constitutive_2000}
\begin{equation}
  \bm{F} = \bm{F}_e\,\bm{F}_p, \qquad \bm{F}_p = \bm{F}{_p}_e\,\bm{F}{_p}_i,
  \label{split_F}
\end{equation}
while $\bm{F}{_p}_e$ denotes the recoverable and $\bm{F}{_p}_i$ the irrecoverable parts of the plastic part of the deformation gradient.\par
\textbf{Helmholtz free energy.}
The Helmholtz free energy is assumed to be a scalar-valued isotropic function and consists of an elastic $\psi_e$ and plastic part $\psi_p$
\begin{equation}
  \psi = \psi_e + \psi_p.
\end{equation} The elastic energy $\psi_e$ is chosen to be of a compressible Neo-Hookean-type and is derived in terms of the volumetric-isochoric split~\cite{steinmann_hyperelastic_2012} of the elastic right Cauchy-Green tensor $\bar{\bm{C}}_e$. The purpose behind this lies in the possibility of being able to compare the obtained results with the element formulation provided by \citet{sauren_mixed_2023, sauren_stability_2024,sauren_mesh_2024}, which depends on this particular split. Appendix~\ref{SBFEM_theorie} describes the relevance of this split. Based on this, the Helmholtz free energy is chosen and expressed in the co-rotated intermediate configuration, denoted with a bar on top $\bar{(\cdot)}$, as follows
\begin{equation}
  \psi_e = \underbrace{\frac{\mu}{2}\biggl(\tr{\bigl((\det{(\bar{\bm{C}}_e}))^{-\frac{1}{3}}\,\bar{\bm{C}}_e\bigr)-3}\biggr)}_{\psi_e^{\mathrm{dev}}} + \underbrace{\frac{\kappa}{2}\biggl(\sqrt{\det{(\bar{\bm{C}}}_e)} - 1\biggr)^2}_{\psi_e^{\mathrm{vol}}}
\end{equation}
with $\mu$ and $\lambda$ being the two Lam\'{e} constants and $\kappa = \lambda + \frac{2}{3}\,\mu$ being the bulk modulus. The isochoric elastic right Cauchy-Green tensor is defined as $\tilde{\bar{\bm{C}}}_e = (\det{(\bar{\bm{C}}_e)})^{-1/3}\,\bar{\bm{C}}_e$, see~\cite{steinmann_hyperelastic_2012}. The plastic energy is defined as
\begin{equation}
  \psi_p = \frac{a}{2}\,(\tr{(\bar{\bm{B}}{_p}_e)} - 3 - \ln{(\det{(\bar{\bm{B}}{_p}_e}))} + e \,\biggl(\xi_p + \frac{\exp{(-f\,\xi_p)} - 1}{f} \biggr)
\end{equation}
with the isotropic hardening variable $\xi_p$ and $\bar{\bm{B}}{_p}_e = \bm{F}{_p}_e\bm{F}{_p}_e^T$. The first term depicts kinematic hardening with the stiffness-like material parameter $a$ while the second term is responsible for modeling nonlinear Voce isotropic hardening with the stiffness-like material parameter $e$ and the dimensionless material parameter $f$~\cite{ voce1955practical}. The quantities on which the Helmholtz free energy depends, are defined in the co-rotated intermediate configuration and thus pulled back as follows 
\begin{align*}
  \bar{\bm{C}}_e := \bm{U}_p^{-1}\bm{C}\,\bm{U}_p^{-1}, \qquad
  \bar{\bm{B}}{_p}_e := \bm{U}_p\,\bm{C}{_p}_i^{-1}\,\bm{U}_p \qquad \text{with:}\,\,\bm{C}{_p}_i = \bm{F}{_p}_i^{T}\bm{F}{_p}_i = \bm{U}{{_p}_i}\,\bm{U}{{_p}_i}.
\end{align*}
Here, $\bm{C} = \bm{F}^T\bm{F}$ denotes the right Cauchy-Green tensor while $\bm{C}{_p}_i$ stems from the additional split, cf. Equation~\eqref{split_F}. The same holds for the plastic part of the stretch tensor $\bm{U}_p$, where $\bm{U}{{_p}_i}$ results from the additional split of the plastic part of the deformation gradient.
For a more detailed explanation, the reader is kindly referred to~\cite{holthusen_inelastic_2023}. Satisfying the second law of thermodynamics, all needed quantities are derived based on the isothermal Clausius--Duhem inequality. Using the Coleman--Noll procedure~\cite{coleman_foundations_1961} yields the following expression for the second Piola-Kirchhoff stress tensor 
\begin{equation}
  \bm{S} = 2\,\bm{U}_p^{-1}\,\frac{\hat{\delta} \psi_e}{\hat{\delta} \bar{\bm{C}}_e}\,\bm{U}_p^{-1},
\end{equation}
where $\frac{\hat{\delta}(\cdot)}{\hat{\delta}(\cdot)}$ denotes the derivative obtained using AD. The volumetric part of the second Piola-Kirchhoff stress tensor and the pressure needed for the mixed scaled boundary finite element formulation of \citet{sauren_mesh_2024} are obtained as follows
\begin{equation}
  \bm{S}^{\mathrm{vol}} = 2\,\bm{U}_p^{-1}\,\frac{\partial \psi_e^{\mathrm{vol}}}{\partial \bar{\bm{C}}_e}\,\bm{U}_p^{-1}, \qquad p = \frac{\partial \psi_e^{\mathrm{vol}}}{\partial J}
  \label{Svol_press}
\end{equation}
with $J = \sqrt{\det{\bm{C}}}$. A more detailed derivation can be found in Appendix~\ref{SBFEM_theorie}.\par
\textbf{Yield function and evolution equations.}
The yield criterion $\Phi$ depends on both kinematic and isotropic hardening. The potential $g_{\mathrm{kin}}$ is suitable for describing the evolution of nonlinear  Armstrong--Frederick kinematic hardening~\cite{armstrong1966mathematical}. Both are defined as follows
\begin{align}
  \Phi &= \sqrt{\frac{3}{2}\,\tr{(\dev{\bar{\bm{\Gamma}}}^2})} - (\sigma_{y0} + q_p),\qquad
  g_{\mathrm{kin}} = \frac{b}{2a} \, \tr{(\dev{\check{\bm{\Theta}}}}^2)
\end{align}
with the hardening force $q_p = \frac{\hat{\delta} \psi}{\hat{\delta} \xi_p}$ and the material parameter $\sigma_{y0}$ representing the initial yield stress. The material parameter $b$ depicts the kinematic hardening. Here, $\check{\bm{\Theta}}$ denotes the driving force in the co-rotated intermediate configuration 
\begin{equation}
  \check{\bm{\Theta}} = 2\, \bm{U}{_p}_i \,\bm{\Theta}\, \bm{U}{_p}_i, \qquad \bm{\Theta} = \bm{C}{_p}_i^{-1}\,\bm{U}_p\,\frac{\hat{\delta} \psi}{\hat{\delta}\bar{\bm{B}}{_p}_e}\, \bm{U}_p\,\bm{C}{_p}_i^{-1}.
\end{equation}
Based on this, the evolution equations are defined as
\begin{equation}
  \bar{\bm{D}}_p = \dot{\gamma}_p\,\frac{\hat{\delta} \Phi}{\hat{\delta} \bar{\bm{\Gamma}}}, \qquad \check{\bm{D}}{_p}_i = \dot{\gamma}_p\, \frac{\hat{\delta} g_{\mathrm{kin}}}{\hat{\delta} \check{\bm{\Theta}}},
\end{equation}
where $\dot{\gamma}_p$ represents the plastic multiplier. The symmetric relative-stress-like quantity in the co-rotated intermediate configuration $\bar{\bm{\Gamma}} = \bar{\bm{\Sigma}} - \bar{\bm{\chi}}$ denotes the difference between the stress-like Mandel tensor $\bar{\bm{\Sigma}}$ and the stress-like Back tensor $\bar{\bm{\chi}}$, which are derived from the Clausius--Duhem inequality and yield 
\begin{equation}
  \bar{\bm{\Sigma}} = 2\, \bar{\bm{C}}_e\, \frac{\hat{\delta} \psi }{\hat{\delta} \bar{\bm{C}}_e}, \qquad \bar{\bm{\chi}} = 2\, \frac{\hat{\delta} \psi}{\hat{\delta} \bar{\bm{B}}{_p}_e}\,\bar{\bm{B}}{_p}_e.
\end{equation}
At last, the Karush-Kuhn-Tucker conditions are taken into consideration
\begin{equation}
    \Phi \leq 0, \qquad\dot{\gamma}_p \geq 0, \qquad \dot{\gamma}_p\,\Phi = 0.
\end{equation}

\section{Two-field variational functional}
\label{Two-field variational functional}
The two-field functional with respect to the reference configuration results from the enhanced assumed strain (EAS) concept by \citet{simo_geometrically_1992,simo_improved_1993}, who based this derivation on the Hu-Washizu variational principle 
\begin{subequations}
\begin{align}
  \label{lin_momen}
  g_u^c(\bm{u},\bm{w}, \delta \bm{u}) &:= \int_{\mathcal{B}_0} \bm{S}(\bm{E})\, : \, \delta \bm{E}_c \, \mathrm{d}V - \int_{\mathcal{B}_0}\, \bm{f}_0 \cdot \delta \bm{u} \,\mathrm{d}V - \int_{\partial_t\mathcal{B}_0} \bm{t}_0 \cdot \delta \bm{u} \, \mathrm{d}A = 0, \\
  \label{orthogonality part}
  g_w(\bm{u},\bm{w}, \delta \bm{u}) &:= \int_{\mathcal{B}_0} \bm{S}(\bm{E})\, : \, \delta \bm{E}_{\mathrm{enh}} \, \mathrm{d}V = 0.
\end{align}
\end{subequations}
Here, Equation~\eqref{lin_momen} comes from the balance of linear momentum while Equation~\eqref{orthogonality part} denotes the orthogonality condition. The compatible displacement field is denoted as $\bm{u}$, while $\bm{w}$ describes the incompatible displacement field. The second Piola-Kirchhoff stress tensor $\bm{S}(\bm{E})$ depends on the total Green-Lagrange strain tensor $\bm{E} = \bm{E}_c(\bm{u})  +\bm{E}_{\mathrm{enh}}(\bm{w}) $, which is assumed to be additively decomposable into a compatible (displacement-based) part $\bm{E}_c(\bm{u})$ and an incompatible (enhanced) part $\bm{E}_{\mathrm{enh}}(\bm{w})$. The same assumption also holds for its variation $\delta \bm{E}$. 

\section{Kinematics}
\label{Kinematics}
The domain $\mathcal{B}_0$ is approximated by $\mathcal{B}_0^h$ and discretized into $\nel$ elements $\mathcal{B}_0 \approx \mathcal{B}_0^h = \bigcup\limits_{e=1}^{\nel}\mathcal{B}{_0}_e$.
For a standard eight-node hexahedral finite element, the isoparametric concept is employed with the (initial) geometry $\bm{x}_e$ and the displacement $\bm{u}_e$ being discretized as follows
\begin{align}
  \bm{x}_e \approx \bm{x}_e^h = \bm{N}(\bm{\xi})\,\bm{X}_e, \qquad \bm{u}_e  \approx \bm{u}_e^h= \bm{N}(\bm{\xi})\,\bm{U}_e.
\end{align}
The element nodal positions and the element nodal displacements are stored in the 24$\times$1 vectors $\bm{X}_e$ and $\bm{U}_e$, respectively. Trilinear shape functions $N_I = 1/8 \, (1+\xi_I \xi)(1+\eta_I \eta)(1+\zeta_I \zeta)$ are used for the reference element and stored in the shape function matrix $\bm{N} = (N_1\,\bm{I}_{3},...\,, N_8\,\bm{I}_3)$, depending on the natural coordinate vector $\bm{\xi} = (\xi, \eta, \zeta)^T$ and the 3$\times$3 identity tensor $\bm{I}_3$. Based on this, the Jacobian matrices with respect to the reference and current domain can be calculated using AD as
\begin{align}
  \bm{J} = \frac{\hat{\delta}\bm{x}_e}{\hat{\delta}\bm{\xi}}, \qquad \bm{J}_{\mathrm{cur}} = \frac{\hat{\delta}(\bm{x}_e + \bm{u}_e)}{\hat{\delta}\bm{\xi}},
  \label{Jacobian}
\end{align}
respectively. This leads to the following definition of the deformation gradient $\bm{F}$
\begin{equation}
  \bm{F} = \bm{J}_{\mathrm{cur}}\,\bm{J}^{-1}.
\end{equation}

\textbf{Compatible strain field.}
\label{Compatible strain field}
The total compatible part of the Green-Lagrange strain tensor in the cartesian basis system is computed as follows
\begin{equation}
  \bm{E}_c = \frac{1}{2}\,(\bm{F}^T\bm{F} - \bm{I}_3).
  \label{compatible strain}
\end{equation}
The compatible B-Operator $\bm{B}_c$ can be obtained by using AD as follows 
\begin{equation}
  \bm{B}_c = \frac{\hat{\delta}\hat{\bm{E}_c}}{\hat{\delta}\bm{U}_e}.
  \label{B_c}
\end{equation}
while $\hat{\bm{E}_c}$ denotes the Green-Lagrange strain in Nye's notation. The same discretization holds for the variational and linearized compatible strains $\delta\hat{\bm{E}_c} = \bm{B}_c \,\delta\bm{U}_e$ and $\Delta\hat{\bm{E}_c} = \bm{B}_c \, \Delta\bm{U}_e$. 

\textbf{Enhanced assumed strain field.}
Locking leads to an underestimation of the deformation and an overestimation of the stresses. While shear locking occurs in bending-dominated problems, volumetric locking is present in the case of nearly incompressible material behavior. This is due to the low-order element formulation (linear interpolation of the displacement field). In case of bending, this low-order formulation is not able to depict the element's curvature sufficiently. Since the bending mode can only be characterized in terms of a trapezoidal deformation, unphysical transverse shear strains $E_{c\,\xi\,\eta}$, $E_{c\,\eta\,\zeta}$ and $E_{c\,\zeta\,\xi}$ occur~\cite{barfusz_single_2021}. To tackle this problem, the EAS concept is used, where the compatible Green-Lagrange strain vector $\hat{\bm{E}_c}$ is enriched by an incompatible (enhanced) part $\hat{\bm{E}}_{\mathrm{enh}}$. Hence, the enhanced B-Operator $\bar{\bm{B}}_{\mathrm{enh}}$ is defined in a way that shear locking is alleviated. Consequently, the discretized incompatible Green-Lagrange strain tensor given in Nye's notation reads
\begin{equation}
  \hat{\bm{E}}_{\mathrm{enh}} = \left[\begin{array}{c}
    0\\
    0  \\
    0\\
    E_{\mathrm{enh}\,\xi \eta}\\
    E_{\mathrm{enh}\,\xi \zeta}  \\
    E_{\mathrm{enh}\,\eta \zeta} 
  \end{array}\right] =  \bm{T}^0\,\left[\begin{array}{cccccc}
    0 & 0 & 0 & 0 & 0 & 0 \\
    0 & 0 & 0 & 0 & 0 & 0 \\
    0 & 0 & 0 & 0 & 0 & 0 \\
    \eta & \xi & 0 & 0 & 0 & 0 \\
    0 & 0 & \zeta & \eta & 0 & 0 \\
    0 & 0 & 0 & 0 & \xi & \zeta 
  \end{array}\right] \,\bm{W}_e = \bm{T}^0\, \bar{\bm{B}}_{\mathrm{enh}} \,\bm{W}_e = \bm{B}_{\mathrm{enh}} \,\bm{W}_e.
  \label{enhanced strain}
\end{equation}
Here, $\bm{W}_e = (W_1, W_2, W_3, W_4, W_5, W_6)^T$ contains the six additional enhanced degrees-of-freedom, see~\cite{schwarze_reduced_2011}, and $\bm{T}^0$ is the transformation matrix evaluated at the center of the element. Analogously to Equation~\eqref{enhanced strain}, its variation is discretized as $\delta\hat{\bm{E}}_{\mathrm{enh}} = \bm{B}_{\mathrm{enh}} \,\delta\bm{W}_e$. In general, the transformation matrix $\bm{T}$ ensures a connection between the cartesian and convective (natural) coordinates. A definition of the transformation matrix is shown in Appendix~\ref{t_matrix}. The introduction of convective (natural) coordinates is a necessity to account for certain different locking phenomena, e.g. curvature thickness locking as well as transverse shear locking. For example, within the scope of the assumed natural strain (ANS) method, certain strain components are modified with respect to their convective coordinates for element formulations such as the solid-shell of \citet{schwarze_reduced_2009, schwarze_reduced_2011} or the solid-beam of \citet{frischkorn_solid-beam_2013}. 
\section{Concept of reduced integration}
\label{Gauss_point_concept}
In order to be able to compute the residual vectors and the corresponding element stiffness matrix for the formulation presented above, the Gaussian quadrature is used. In case of an eight-node hexahedral element with trilinear shape functions at least eight integration points are needed in order to obtain an accurate result, which can be computationally costly. Thus, the idea of reduced integration was introduced. However, just using reduced integration can lead to a rank-deficient stiffness matrix and usually causes the element response to be too softly. To remedy this, hourglass stabilization techniques are used. In particular, a Taylor series expansion up to the bilinear terms of all needed quantities is applied in order to be able to integrate the hourglass terms analytically and on top of that achieve a computationally efficient element formulation.\par
\textbf{Taylor series expansion of the constitutively dependent quantities.}
A Taylor series expansion of the constitutively dependent quantities with respect to the center of the element enables the possibility of separating the weak form into physically reasonable parts $(\cdot)^0$, and hourglass parts $(\cdot)^{hg}$. The purpose of the hourglass parts lies in stabilizing and thus avoiding possible rank deficiency in the element formulation. Furthermore, since these parts are represented by means of polynomials, analytical integration of the hourglass parts becomes possible. A Taylor series expansion of the second Piola-Kirchhoff stress in Nye's notation $\hat{\bm{S}}$ up to the bilinear terms is carried out
\begin{align}
  \begin{split}
  \hat{\bm{S}}(\hat{\bm{E}}) &\approx \hat{\bm{S}}\,\bigg|_{\bm{\xi}=\bm{0}} + \bm{C}^{hg} \Biggl( \sum_{i = 1}^{3} \frac{\hat{\delta}\hat{\bm{E}}}{\hat{\delta}\xi_i}\,\bigg|_{\bm{\xi}=\bm{0}}\xi_i + \frac{1}{2}\,\sum_{i = 1}^{3} \sum_{\substack{j=1 \\ j\neq i}}^{3} \biggl( \frac{\hat{\delta}}{\hat{\delta}\xi_j} \biggl( \frac{\hat{\delta}\hat{\bm{E}}}{\hat{\delta}\xi_i}\biggr)\biggr)\,\bigg|_{\bm{\xi}=\bm{0}}\xi_i\,\xi_j\Biggr)\\
  &= \hat{\bm{S}}^0 + \bm{C}^{hg} \Biggl(\underbrace{\hat{\bm{E}}^{\spacee\xi}_c \, \xi + \hat{\bm{E}}^{\spacee\eta}_c \, \eta + \hat{\bm{E}}^{\spacee\zeta}_c  \zeta}_{=: \hat{\bm{E}}^{\mathrm{hg1}}_c} + \underbrace{\hat{\bm{E}}^{\spacee\xi \eta}_c\, \xi  \eta  + \hat{\bm{E}}^{\spacee\xi \zeta}_c\, \xi \zeta + \hat{\bm{E}}^{\spacee\eta \zeta}_c\, \eta \zeta}_{=: \hat{\bm{E}}^{\mathrm{hg2}}_c} + \underbrace{\hat{\bm{E}}^{\spacee\xi}_{\mathrm{enh}}\xi + \hat{\bm{E}}^{\spacee\eta}_{\mathrm{enh}}\eta + \hat{\bm{E}}^{\spacee\zeta}_{\mathrm{enh}}\zeta}_{= \hat{\bm{E}}_{\mathrm{enh}}}\Biggr)\\
  &= \hat{\bm{S}}^0 +  \underbrace{\bm{C}^{hg}(\hat{\bm{E}}^{\mathrm{hg1}}_c + \hat{\bm{E}}^{\mathrm{hg2}}_c + \hat{\bm{E}}_{\mathrm{enh}})}_{=: \hat{\bm{S}}^{\mathrm{hg}}}.
  \label{SPK_TE}
  \end{split}
\end{align}
In this context, the superscript denotes the derivative with respect to the convective coordinates. Based on this, a polynomial form of both $\bm{B}_c$ and $\bm{B}_{\mathrm{enh}}$ is obtained, see Appendix~\ref{res_stiff} for a more detailed derivation. The so-called hourglass tangent $\bm{C}^{hg}$ is defined based on a linear-elastic material behavior. It is defined in such a way that $\hat{\bm{S}}^{\mathrm{hg}}$ has deviatoric character in order to overcome volumetric locking, see~\cite{schwarze_reduced_2011}
\begin{align}
  \bm{C}^{hg} =  \frac{\mu^{hg}_{\mathrm{eff}}}{3} \left[\begin{array}{cccccc}
    4 & -2 & -2 & 0 & 0 & 0 \\
    -2 & 4 & -2 & 0 & 0 & 0 \\
    -2 & -2 & 4 & 0 & 0 & 0 \\
    0 & 0 & 0 & 3 & 0 & 0 \\
    0 & 0 & 0 & 0 & 3 & 0 \\
    0 & 0 & 0 & 0 & 0 & 3 
  \end{array}\right], \quad \mu^{hg}_{\mathrm{eff}} = \frac{1}{2} \sqrt{\frac{\tr{\Bigl(\dev{\bm{S}^0}^2\Bigr)}}{\tr{\Bigl(\dev{\bm{E}^0_c}^2}\Bigr)}}.
\end{align}
The quantity $\mu^{hg}_{\mathrm{eff}}$ defines the so-called effective shear modulus. In correspondence to \citet{barfusz_single_2021}, $\mu^{hg}_{\mathrm{eff}}$ is stored as a history variable and taken from the last converged step in order to circumvent the need for linearization. The use of $\bm{C}^{hg}$ instead of the nonlinear material tangent that is obtained based on the choice of the material model, which would usually result from the derivation of $\hat{\bm{S}}$ with respect to $\hat{\bm{E}}$, avoids the necessity of linearization and enables analytical integration of the residual vectors and thus the stiffness matrix~\cite{barfusz_single_2021}. With the use of AD, all presented derivatives can be directly computed, ensuring a relatively simple element implementation. Hence, they do not need to be derived by hand, which would usually be time-consuming and error-prone. Especially quantities, e.g. the compatible B-Operator $\bm{B}_c$ do not have to be directly derived since it can be computed by the relation of the compatible Green-Lagrange strain, see Equations~\eqref{compatible strain} and~\eqref{B_c}.

\section{Enhanced Q1STc by an approximation-free computation of the inverse Jacobian}
\label{Approx_Jaci}
For elements with arbitrary geometries, an analytical integration is not possible, since the inverse Jacobian yields highly nonlinear results. Previous works of \citet{reese_large_2007} and \citet{legay_elastoplastic_2003} introduced the idea of evaluating the inverse Jacobian at the center of the element and thus enabling analytical integration. However, the underlying assumption that the Jacobian would then be constant is only true for very fine meshes. It is stated in \citet{schwarze_reduced_2011, schwarze_reduced_2009} that a linear approximation of the inverse Jacobian yields more realistic results, especially if the element does not correspond to an exact parallelepiped shape. Hence, a Taylor series expansion up to the linear terms is carried out for the inverse Jacobian $\bm{J}^{-1}$ 
\begin{equation}
  \bm{J}^{-1} \approx \bm{J}^{-1} \,\bigg|_{\bm{\xi}=\bm{0}} + \sum_{i = 1}^{3} \frac{\hat{\delta}\bm{J}^{-1}}{\hat{\delta}\xi_i}\,\bigg|_{\bm{\xi}=\bm{0}}\xi_i.
  \label{TE Jac Inv}
\end{equation}

Taking the Taylor expansion of $\bm{J}\bm{J}^{-1}$ and considering that $\bm{J}\bm{J}^{-1} = \bm{I}$, which must be satisfied for any arbitrary chosen natural coordinate vector $\bm{\xi}$, yields an expression for the partial derivation of the linear part in Equation~\eqref{TE Jac Inv} 
\begin{equation}
  \frac{\hat{\delta}\bm{J}^{-1}}{\hat{\delta}\xi_i}\,\bigg|_{\bm{\xi}=\bm{0}} = -(\bm{J}^{0})^{-1}\bm{J}^{\xi_i}(\bm{J}^{0})^{-1}.
  \label{Jinv_frag}
\end{equation}
With Equations~\eqref{TE Jac Inv} and~\eqref{Jinv_frag}, the inverse Jacobian reads
\begin{equation}
  \bm{J}^{-1} \approx \bm{J}^{-1} \,\bigg|_{\bm{\xi}=\bm{0}} - \sum_{i = 1}^{3}(\bm{J}^{0})^{-1}\bm{J}^{\xi_i}(\bm{J}^{0})^{-1}\xi_i. 
  \label{TE_Inv_Jac}
\end{equation}
This leads to the following polynomial representation for the inverse Jacobian with $\bm{J}^{-1} = \bm{j}$
\begin{equation}
  \bm{j} \approx \bm{j}^0 + \bm{j}^{\spacee\xi} \, \xi + \bm{j}^{\spacee\eta} \, \eta + \bm{j}^{\spacee\zeta} \, \zeta.
  \label{j_Taylor}
\end{equation}
For a more detailed derivation, the reader is kindly referred to~\cite{schwarze_reduced_2011}. This approach allows for a polynomial approximation of the transformation matrix $\bm{T}$
\begin{equation}
  \bm{T} \approx \bm{T}^0 + \bm{T}^{\spacee\xi} \, \xi + \bm{T}^{\spacee\eta} \, \eta + \bm{T}^{\spacee\zeta} \, \zeta.
  \label{TE_Tmatrix}
\end{equation}
With the use of AD, an approximation of the inverse Jacobian using a Taylor series expansion becomes redundant. Hence, the Q1STc formulation is enhanced by an approximation-free computation of the inverse Jacobian, termed 'Q1STc+'. 
The following algorithms demonstrate the use of AD for computing the inverse Jacobian. Algorithm~\ref{Q1STc_pseudocode} shows the steps for the computation of an approximated inverse Jacobian by a Taylor series expansion, which can be done by hand, while Algorithm~\ref{Q1STcpl_pseudocode} demonstrates a direct calculation. A direct computation of the inverse Jacobian leads to a more simplified code, as the process is reduced to a single step. It needs to be taken into account that with a derivation purely by hand without AD, a direct computation of $\bm{j}$ would cause difficulties for the derivation of $\hat{\bm{E}}_c$ by a Taylor series approximation, cf. Equation~\eqref{SPK_TE}, since this quantity depends on the highly nonlinear inverse Jacobian $\bm{j}$.

\begin{minipage}[t]{0.48\textwidth}
  \floatname{algorithm}{Algorithm}
  \begin{algorithm}[H]
    \caption{Excerpt for the computation of the inverse Jacobian for Q1STc}\label{Q1STc_pseudocode}
    \begin{algorithmic}
      \LineComment {\text{Compute Jacobian}}
      \vspace{0.3em}
      \State $\bm{J} \gets \frac{\partial\,(\bm{N}(\bm{\xi})\,\bm{X}_e)}{\partial\bm{\xi}}$
      \LineComment {\text{Taylor series expansion of the Jacobian}\par 
      (Computable by hand)} 
      \vspace{0.3em}
      \State $\bm{J}^0 \gets \bm{J}\,\big|_{\bm{\xi}=\bm{0}}$
      \vspace{0.1em}
      \State $\bm{J}^{\xi} \gets \frac{\partial\bm{J}}{\partial\xi}\,\big|_{\bm{\xi}=\bm{0}}$
      \vspace{0.1em}
      \State $\bm{J}^{\eta} \gets \frac{\partial\bm{J}}{\partial\eta}\,\big|_{\bm{\xi}=\bm{0}}$
      \vspace{0.1em}
      \State $\bm{J}^{\zeta} \gets \frac{\partial\bm{J}}{\partial\zeta}\,\big|_{\bm{\xi}=\bm{0}}$
      \vspace{0.3em}
      \LineComment {\text{Approximation of the inverse Jacobian}}
      \vspace{0.3em}
      \State $\bm{j}^{0}\gets (\bm{J}^0)^{-1}$
      \vspace{0.1em}
      \State $\bm{j}^{\xi}\gets -\bm{j}^{0}\bm{J}^{\xi}\bm{j}^{0}$
      \vspace{0.1em}
      \State $\bm{j}^{\eta}\gets -\bm{j}^{0}\bm{J}^{\eta}\bm{j}^{0}$
      \vspace{0.1em}
      \State $\bm{j}^{\zeta}\gets -\bm{j}^{0}\bm{J}^{\zeta}\bm{j}^{0}$
      \vspace{0.3em}
      \LineComment {\text{Inverse Jacobian}}
      \vspace{0.3em}
      \State $\bm{j} \gets \bm{j}^{0} + \bm{j}^{\xi}\, \xi + \bm{j}^{\eta}\,\eta + \bm{j}^{\zeta}\, \zeta$
    \end{algorithmic}
  \end{algorithm}
  \end{minipage}
  \hfill
  \begin{minipage}[t]{0.48\textwidth}
  \floatname{algorithm}{Algorithm}
  \begin{algorithm}[H]
    \caption{Excerpt for the computation of inverse Jacobian for Q1STc+}\label{Q1STcpl_pseudocode}
    \begin{algorithmic}
      \LineComment{\text{Compute Jacobian}}
      \vspace{0.3em}
      \State $\bm{J} \gets \frac{\hat{\delta}(\bm{N}(\bm{\xi})\,\bm{X}_e)}{\hat{\delta}\bm{\xi}}$
      \vspace{0.3em}
      \LineComment{\text{Inverse Jacobian}}
      \vspace{0.3em}
      \State $\bm{j} \gets (\bm{J})^{-1}$
    \end{algorithmic}
  \end{algorithm}
  \end{minipage}

\section{Discretization of the weak form}
\label{Discretization of the weak form}
The first part of the two-field variational relation in~\eqref{lin_momen} needs to be assembled on a global finite element level $\bigl(g_u^c = \bigcup\limits_{e=1}^{\nel} g_{u_e}^c\bigr)$. The second variational equation~\eqref{orthogonality part} has to be solved at the element level. Hence, equation~\eqref{orthogonality part} yields with~\eqref{enhanced strain} and~\eqref{SPK_TE} 
\begin{align}
  \begin{split}
    g_{w_e} &= \delta \bm{W}_e^T  \underbrace{\int_{V_e} \bm{B}_{\mathrm{enh}}^T \hat{\bm{S}}^0\,\mathrm{d}V_e^0}_{=\bm{0}} \,+ \, \delta \bm{W}_e^T  \underbrace{\int_{V_e} \bm{B}_{\mathrm{enh}}^T \bm{C}^{\mathrm{hg}} \hat{\bm{E}}_c^{\mathrm{hg2}}\,\mathrm{d}V_e^0}_{=\bm{0}}\\ &+ \,\delta \bm{W}_e^T \underbrace{\int_{V_e} \bm{B}_{\mathrm{enh}}^T \bm{C}^{\mathrm{hg}} \hat{\bm{E}}_c^{\mathrm{hg1}}\,\mathrm{d}V_e^0}_{=: \bm{R}_{w_e}} \,+ \,\delta \bm{W}_e^T \underbrace{\int_{V_e} \bm{B}_{\mathrm{enh}}^T \bm{C}^{\mathrm{hg}} \bm{B}_{\mathrm{enh}}\,\mathrm{d}V_e^0}_{=: \bm{K}_{ww_e}} \,\bm{W}_e \overset{!}{=}0.
    \label{g_w_dis}
  \end{split}
\end{align}
Here, the volume element is approximated by $\mathrm{d}V_e \approx \mathrm{d}V_e^0 = \det{(\bm{J}^0)}\,\mathrm{d}\xi\,\mathrm{d}\eta\, \mathrm{d}\zeta$ with $\bm{J}^0$ denoting the Jacobian evaluated at the center. Based on~\eqref{g_w_dis}, the vector $\bm{W_e}$ can be expressed in terms of $\bm{U}_e$ as follows
\begin{equation}
  \bm{W}_e = -(\bm{K}_{ww_e})^{-1} \bm{R}_{w_e}(\bm{U}_e).
  \label{W_esolution}
\end{equation}
The relation for $\bm{W_e}$ in~\eqref{W_esolution} is then inserted into Equation~\eqref{lin_momen} in Nye's notation, reducing the two-field variational functional into a primal formulation. Due to this, $\bm{W_e}$ is eliminated at element level and does not have to be determined explicitly. Considering only the internal work of the first variational function in~\eqref{lin_momen} and using Equations~\eqref{compatible strain},~\eqref{B_c},~\eqref{enhanced strain} and~\eqref{SPK_TE} yields 
\begin{align}
  \begin{split}
    g_{u_e}^c &= \delta \bm{U}_e^T  \underbrace{\int_{V_e} (\bm{B}^0_c)^T \hat{\bm{S}}^0 \,\mathrm{d}V_e^0}_{=:\bm{R}_{u_e}^0} \, + \, \delta \bm{U}_e^T  \underbrace{\int_{V_e} (\bm{B}^{\mathrm{hg1}}_c)^T \bm{C}^{\mathrm{hg}}\hat{\bm{B}}_{\mathrm{enh}} \,\mathrm{d}V_e^0}_{=:\bm{K}_{uw_e}}-(\bm{K}_{ww_e})^{-1} \bm{R}_{w_e}\\
    &+ \,\delta \bm{U}_e^T \Biggl[  \underbrace{\int_{V_e} (\bm{B}^{\mathrm{hg1}}_c)^T \bm{C}^{\mathrm{hg}}\hat{\bm{E}}^{\mathrm{hg1}}_c \,\mathrm{d}V_e^0
    \,+\, \int_{V_e} (\bm{B}^{\mathrm{hg2}}_c)^T \bm{C}^{\mathrm{hg}}\hat{\bm{E}}^{\mathrm{hg2}}_c \,\mathrm{d}V_e^0}_{=:\bm{R}_{u_e}^{\mathrm{hg}}} \Biggr],
    \label{g_u_dis}
  \end{split}
\end{align}
 A more detailed derivation leading to these expressions can be found in \citet{barfusz_single_2021}. The residuals $\bm{R}_{w_e}$, $\bm{R}_{u_e}^0$ and $\bm{R}^{\mathrm{hg}}_{u_e}$ and the matrices $\bm{K}_{ww_e}$  and $\bm{K}_{uw_e}$ obtained from~\eqref{g_w_dis} and~\eqref{g_u_dis} can be solved using analytical integration, see for further explanation Section~\ref{Gauss_point_concept} and Appendix~\ref{res_stiff}.

At the end, the internal weak form of~\eqref{g_u_dis} reads 
\begin{equation}
  g_{u_e}^c = \delta \bm{U}_e^T(\bm{R}^0_{u_e} + \bm{R}^{\mathrm{hg}}_{u_e} - \bm{K}_{uw_e}(\bm{K}_{ww_e})^{-1} \bm{R}_{w_e}).
  \label{total_res_wf}
\end{equation}
\textbf{Global assembly.} The Q1STc element formulation of \citet{barfusz_single_2021} and the modified Q1STc+ formulation have been implemented into the FE software \textit{FEAP}~\cite{taylor_feap_nodate}, while the element routines were implemented with help of the automatic differentiation tool \textit{AceGen}~\cite{korelc_multi-language_nodate, korelc2016automation}. The total residual vector at element level reads
\begin{equation}
  \bm{G}_{u_e}(\bm{U}_e) = \bm{R}^0_{u_e}(\bm{U}_e) + \bm{R}^{\mathrm{hg}}_{u_e}(\bm{U}_e) - \bm{K}_{uw_e}(\bm{U}_e)(\bm{K}_{ww_e})^{-1} \bm{R}_{w_e}(\bm{U}_e) - \bm{F}_{\mathrm{ext}_e}
  \label{total_res}
\end{equation}
with $\bm{F}_{\mathrm{ext}_e}$ denoting the external force vector at element level. Since the residual vector is in general nonlinear, an iterative solution scheme is used. Here, the Newton-Raphson scheme is applied, where Equation~\eqref{total_res} is expanded into a Taylor series up to the linear terms
\begin{align}
  \begin{split}
    \bm{G}_{u_e}(\bm{U}_e) &\approx   \tilde{\bm{G}}_{u_e} + \underbrace{\frac{\partial\,\bm{G}_{u_e}}{\partial \,\bm{U}_e}\bigg|_{\tilde{\bm{U}}_e}}_{\tilde{\bm{K}}_e} \,\Delta \bm{U}_e\\
    &= \tilde{\bm{G}}{_u}_e + \tilde{\bm{K}}_e \,\Delta \bm{U}_e,
  \end{split}
\end{align} 
where $\tilde{(\cdot)}$ denotes the evaluation at the current state of $\bm{U}_e$. The stiffness matrix at element level is obtained with help of AD by the following relation
\begin{equation}
  \bm{K}_e = \frac{\hat{\delta}\bm{R}_e}{\hat{\delta}\bm{U}_e}.
  \label{stiffness_total}
\end{equation}

\section{Numerical examples}
\label{Numerical_ex}
The examples at hand serve to test and compare the performance of the Q1STc+ element formulation with the Q1STc formulation. As previously mentioned, the main difference between the Q1STc and Q1STc+ formulation lies in the approximation of the inverse Jacobian and as a result, the transformation matrix, cf. Equations~\eqref{TE Jac Inv} and~\eqref{TE_Tmatrix}. The main purpose of the following examples lies in testing and investigating the influence of this modification, with a particular focus on distorted meshes. To validate the results, a volumetric locking-free polygonal finite element formulation proposed by \citet{sauren_mesh_2024} is used. Noteworthy, in this publication, the formulation is extended to finite elasto-plasticity and denoted as "U-P-SBFEM". The abbreviation "U-P" stands for the mixed displacement-pressure approach while "SBFEM" emphasizes that a scaled boundary finite element formulation is used. In addition, the conventional low-order finite element formulation with full integration using eight Gauss points, e.g.~\citet{zienkiewicz_finite_2010}, and here termed "Q1", is utilized for comparison, as both Q1STc and Q1STc+ are based on this formulation. 
\subsection{Patch tests}
\label{patch_tests}
The purpose of patch tests lies in testing the accuracy of a finite element formulation. Hence, \citet{macneal_proposed_1985} developed certain patch tests which are applicable to any finite element formulation and focus on different kinds of structures, e.g. beams, shells or brick elements. They state that if an element passes the patch test, the outcomes from any other structural example will reach the correct solution. In the following, the membrane patch test and the patch test for solids will be carried out. The focus here lies in testing and comparing the performance of both Q1STc and Q1STc+. For a better comparison, and given that its results align with the analytical solution, satisfying both patch tests, the performance of the conventional low-order finite element formulation with full integration using eight Gauss points (Q1) is used as an additional reference to the computed analytical results.\par
\textbf{Membrane patch test.} The geometry of the patch of elements with a thickness of $t = \text{0.001}\, \text{mm}$ is shown in Figure~\ref{membrane_Struc}. It needs to be taken into consideration that the following figure only shows the first eight nodes at $z = \text{0}$. The remaining eight nodes have the same $x$ and $y$ coordinates but a different $z$ coordinate. The analytical results in~\cite{macneal_proposed_1985} are given for the case of linear elasticity. The given displacements (in $\text{mm}$) are prescribed at the outer nodes of the structure while $u_z = \text{0}$ for the bottom layer of the nodes, resulting in a plane stress state. The prescribed displacements are defined according to the function
\begin{equation}
    \bm{u} = \left(\begin{array}{c} u_x \\ u_y\end{array}\right) = \left(\begin{array}{c} 0.001\,(x+0.5y) \\ 0.001\,(0.5x+y) \end{array}\right).
\end{equation}
Based on using the St.Venant Kirchhoff material model, the analytical second Piola-Kirchhoff stress yields $S_{xx} = S_{yy} = \text{1334.20}\,\text{N}/\text{mm}^2$ for each element in the patch, while the corresponding shear component is equal to $S_{xy} = \text{400.50}\,\text{N}/\text{mm}^2$. The material parameters are chosen according to \citet{macneal_proposed_1985} as $\lambda = \text{400000} \, \text{N}/\text{mm}^2$ and $\mu = \text{400000} \, \text{N}/\text{mm}^2$ $(E = \text{10}^\text{6} \,\text{N}/\text{mm}^2, \nu = \text{0.25})$. It is of importance to mention that neither the stress in $z-$direction nor the shear components related to it should yield stresses since a plane stress state is considered.\par
\begin{figure}[H]
    \centering
    \def\svgwidth{1\textwidth}
    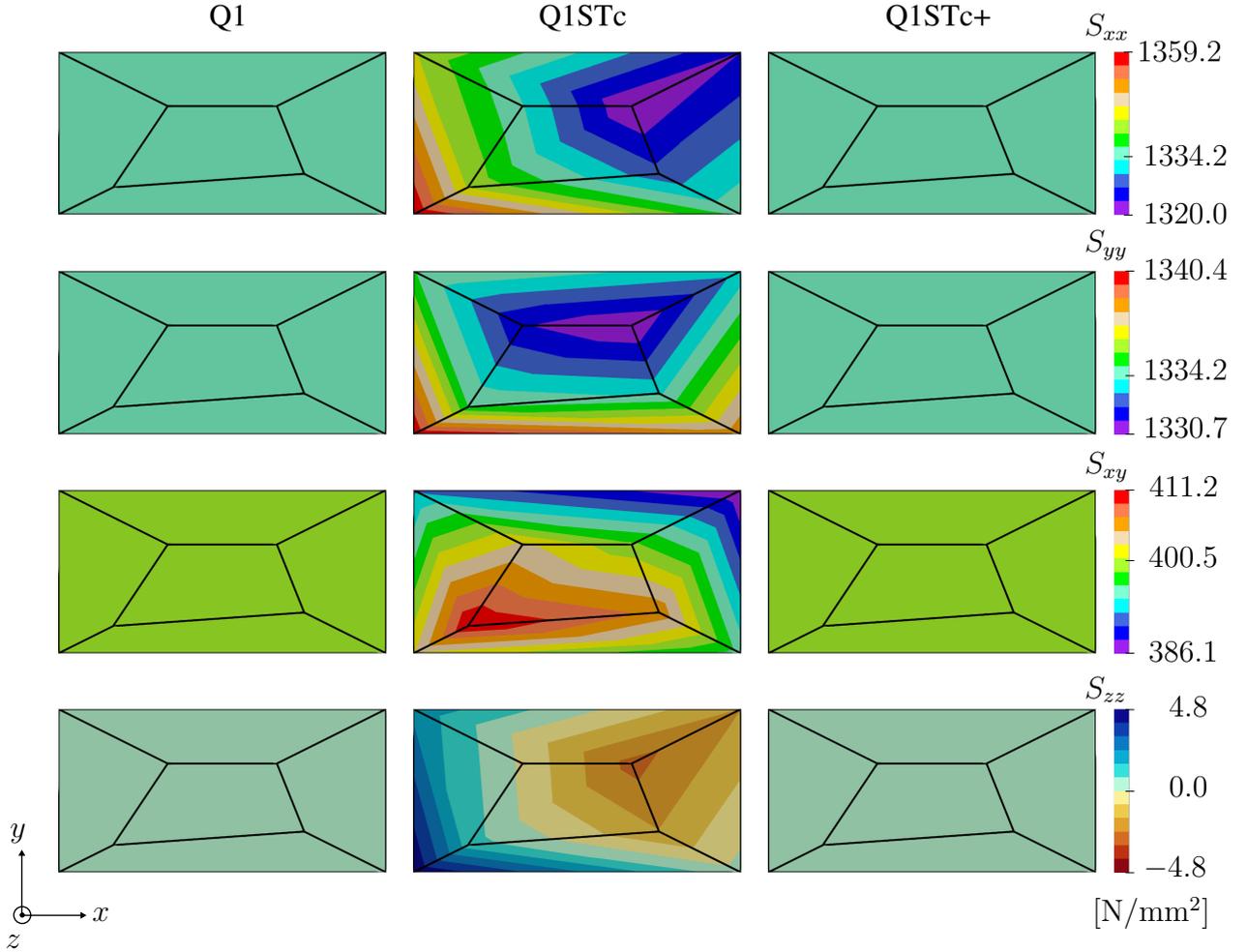
    \caption{Membrane patch test - Stress distribution for Q1, Q1STc and Q1STc+, where Q1 denotes the low-order finite element formulation using full integration, and serves as an additional reference solution to the analytical solution in Table~\ref{membrane_table}.}
    \label{membrane_stresses}
\end{figure}

Table~\ref{membrane_table} shows an overview of the analytical results and the performance of the two element formulations, Q1STc and Q1Stc+. Figure~\ref{membrane_stresses} depicts the distribution of the stresses for both Q1STc and Q1STc+ for each element in the patch while Q1 is used as a reference solution. The patch test can be considered passed if the stresses in the entire structure are homogeneous. However, this is not the case for Q1STc, where the geometry of the element is approximated by a Taylor series expansion, see Equations~\eqref{TE_Inv_Jac} and~\eqref{TE_Tmatrix}. It can be seen that the results for the stress components in $x-$ and $y-$ direction deviate for each element, introducing an error in the range of $\text{0.09}\,\text{\%}- \text{3.3}\,\text{\%}$. Moreover, stresses in $z-$direction are present. Overall, the Q1STc element formulation does not satisfy the membrane patch test. This is due to the fact that the inverse Jacobian and the transformation matrix are approximated by a Taylor series expansion up to the linear terms.  
\begin{figure}[htb]
	\centering
	\input{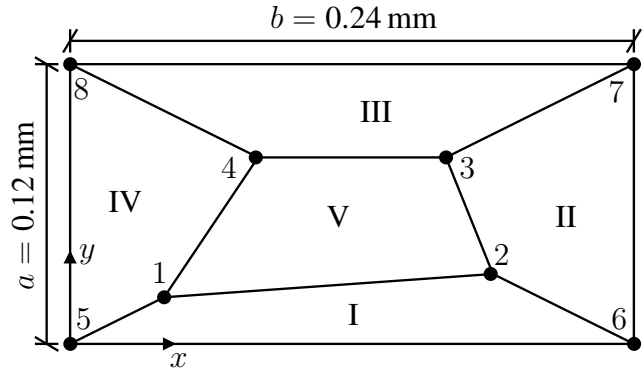}
	\caption{Membrane patch test - Geometry of the five-element patch test with $t = \text{0.001}\, \text{mm}$ with element numbers I - V.}
	\label{membrane_Struc}
\end{figure}
\begin{table}[H]
	\centering
	\caption{Membrane patch test - Stress comparison between the analytical solution and results obtained using Q1STc and Q1STc+, see Figure~\ref{membrane_Struc}.}
	\begin{tabular}{ c c c c c}
	\hline
        $\text{Stress [$\text{N}/\text{mm}^2$]}$ &	$\text{Element number}$	 & $\text{Analytical solution}$ & $\text{Q1STc}$ & $\text{Q1STc+}$ \\
	\hline
	\hline
        \multirow{5}{2em}{$S_{xx}$}	&	I	&	1334.20&	1338.90 & 1334.20\\
        &	II	&	1334.20&	1332.90 & 1334.20\\
        &	III	&	1334.20&	1314.20 & 1334.20\\
        &	IV	&	1334.20&	1376.60 & 1334.20\\
        &	V	&	1334.20&	1313.30 & 1334.20\\
    \hline
        \multirow{5}{2em}{$S_{yy}$}	&	I	&	1334.20&	1335.50 & 1334.20\\
        &	II	&	1334.20&	1341.10 & 1334.20\\
        &	III	&	1334.20&	1326.30 & 1334.20\\
        &	IV	&	1334.20&	1344.60 & 1334.20\\
        &	V	&	1334.20&	1325.00 & 1334.20\\
    \hline
    \multirow{5}{2em}{$S_{zz}$}	&	I	&	0.00&	-2.1025 &0.00\\
    &	II	&	0.00&	2.1957  & 0.00\\
    &	III	&	0.00&	-5.1731 & 0.00\\
    &	IV	&	0.00&	 10.6170 & 0.00\\
    &	V	&	0.00&	 -4.9051 & 0.00\\
	\hline	
    \multirow{5}{2em}{$S_{xy}$}	&	I	&	400.50&	403.75&400.50\\
    &	II	&	400.50&	389.11  & 400.50\\
    &	III	&	400.50&	383.35 & 400.50\\
    &	IV	&	400.50&	 401.50 & 400.50\\
    &	V	&	400.50&	 426.17& 400.50\\
	\hline
    \multirow{5}{2em}{$S_{xz}$}	&	I	&	0.00&	0.0389 &0.00\\
    &	II	&	0.00&	-0.2729  & 0.00\\
    &	III	&	0.00&	0.0007& 0.00\\
    &	IV	&	0.00&	 0.0543 & 0.00\\
    &	V	&	0.00&	0.1197& 0.00\\
	\hline
    \multirow{5}{2em}{$S_{yz}$}	&	I	&	0.00&	-0.35487 &0.00\\
    &	II	&	0.00&	  -0.05452  & 0.00\\
    &	III	&	0.00&	-0.32121& 0.00\\
    &	IV	&	0.00&	 0.09283 & 0.00\\
    &	V	&	0.00&	  0.22888& 0.00\\
	\hline
	\end{tabular}
    \label{membrane_table}
	\end{table}
\par
To test and compare the performance of Q1STc with Q1STc+ more thoroughly, the same test is conducted while an elasto-plastic material is considered. To this end, the theory described in Section~\ref{plasti_theory} is employed. Table~\ref{mat_param_plast} gives an overview of the chosen material parameters. Since a volume-preserving plasticity model is considered, the volumetric locking-free element formulation, extended to elasto-plasticity (U-P-SBFEM), is taken for comparison. \par Figure~\ref{membrane_plasticity} depicts the force-displacement curves of all three element formulations. Here, the displacement of node 7 is plotted versus the sum of the nodal reaction forces at $x = \text{0.24}\,\text{mm}$. For a simpler illustration, the curves are normalized by $F_0 = \text{0.028914}\,\text{N}$, where $F_0$ denotes the largest force obtained by the Q1STc+ formulation in the elastic regime. It can be seen that the proposed Q1STc+ element coincides with U-P-SBFEM. The results of Q1STc, however, yield nodal reaction forces that are severely overestimated, especially when plasticity occurs. The reason for this lies in the approximation of the element geometry, since this is the only difference between the Q1STc and Q1STc+ elements.

\begin{table}[H]
		\centering
		\caption{Membrane patch test -} Material parameters for an elasto-plastic material, see Figure~\ref{membrane_Struc}.
	\begin{tabular}{ l l l l}
	\hline
						$\text{Symbol}$ &	$\text{Material parameter}$	 & $\text{Value}$& $\text{Unit}$	\\
	\hline
	\hline
		$\lambda$ & First Lam\'{e} parameter & 	25000.0 & $\text{N}/\text{mm}^2$\\
		$\mu$	&	 Second Lam\'{e} parameter& 55000.0	&$\text{N}/\text{mm}^2$\\
		$a$	&	 First kinematic hardening stiffness parameter &62.5	& $\text{N}/\text{mm}^2$\\
		$b$	&	Second kinematic hardening stiffness parameter& 2.5	& -\\
		$e$	&	 First isotropic hardening parameter & 125.0	&$\text{N}/\text{mm}^2$\\
		$f$	&	 Second isotropic hardening parameter  &6.0	& -\\
		$\sigma_{y0}$		& Initial plastic threshold & 300.0 &$\text{N}/\text{mm}^2$\\
	\hline	
    \label{mat_param_plast}
	\end{tabular}
\end{table}
\pgfplotsset{%
    width=0.5\textwidth,
    height=0.4\textwidth
}
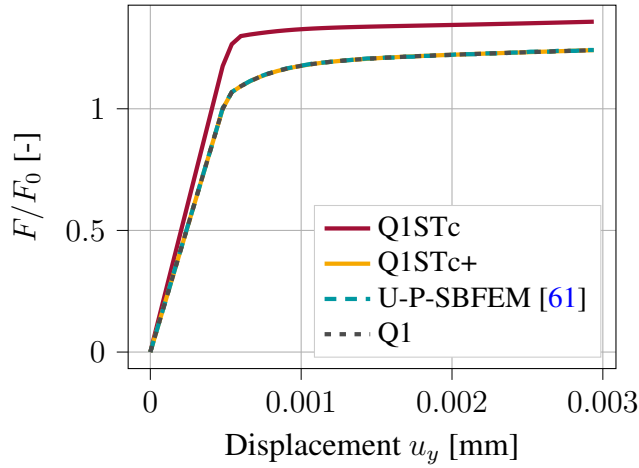
\begin{figure}[H]
    \centering
\begin{tikzpicture}

\definecolor{brown1611653}{RGB}{161,16,53}
\definecolor{darkcyan0152161}{RGB}{0,152,161}
\definecolor{darkgray176}{RGB}{176,176,176}
\definecolor{darkslategray76}{RGB}{76,76,76}
\definecolor{lightgray204}{RGB}{204,204,204}
\definecolor{orange2461680}{RGB}{246,168,0}

\begin{axis}[
legend cell align={left},
legend style={
  fill opacity=1,
  draw opacity=1,
  text opacity=1,
  at={(0.97,0.03)},
  anchor=south east,
  draw=lightgray204,
  font=\fontsize{11}{9}\selectfont,
},
tick align=outside,
tick pos=left,
x grid style={darkgray176},
xlabel={Displacement \(\displaystyle u_y\) [$\text{mm}$]},
xmajorgrids,
xmin=-0.000147, xmax=0.003087,
xtick style={color=black},
y grid style={darkgray176},
ylabel={\(\displaystyle F/F_0\) [$\text{-}$]},
ymajorgrids,
ymin=-0.0678996333955869, ymax=1.42589230130733,
yticklabel style={
        /pgf/number format/fixed,
        /pgf/number format/precision=2
},
xticklabel style={
        /pgf/number format/fixed,
        /pgf/number format/precision=4
},
scaled y ticks=false,
scaled x ticks=false,
xtick={0, 0.001, 0.002, 0.003},
ytick style={color=black}
]
\addplot [ultra thick, brown1611653]
table {%
0 0
6e-05 0.147160545064675
0.00012 0.294272670678564
0.00018 0.441343293906066
0.00024 0.588330912360794
0.0003 0.735283945493533
0.00036 0.882202393304282
0.00042 1.02905167047105
0.00048 1.17586636231583
0.00054 1.26554610223421
0.0006 1.29888635263194
0.00066 1.30521546655599
0.00072 1.31033409421042
0.00078 1.31496852735699
0.00084 1.31904959535173
0.0009 1.32247354222868
0.00096 1.32537870927578
0.00102 1.32783426713703
0.00108 1.3298748011344
0.00114 1.33160406723387
0.0012 1.33309123607941
0.00126 1.33437089299301
0.00132 1.33547762329667
0.00138 1.33648059763436
0.00144 1.33741440132808
0.0015 1.33827903437781
0.00156 1.33907449678357
0.00162 1.33983537386733
0.00168 1.3405962509511
0.00174 1.34135712803486
0.0018 1.34208341979664
0.00186 1.3428442968804
0.00192 1.34357058864218
0.00198 1.34433146572595
0.00204 1.34509234280971
0.0021 1.34585321989348
0.00216 1.34664868229923
0.00222 1.34744414470499
0.00228 1.34827419243273
0.00234 1.34910424016048
0.0024 1.34993428788822
0.00246 1.35076433561596
0.00252 1.3516289686657
0.00258 1.35252818703742
0.00264 1.35339282008716
0.0027 1.35429203845888
0.00276 1.35522584215259
0.00282 1.35612506052431
0.00288 1.35705886421803
0.00294 1.35799266791174
};
\addlegendentry{Q1STc}
\addplot [ultra thick, orange2461680]
table {%
0 7.83876322888566e-14
6e-05 0.125105485232068
0.00012 0.250183302206544
0.00018 0.375216158262433
0.00024 0.500242097253925
0.0003 0.625233450923428
0.00036 0.750190219270941
0.00042 0.875112402296465
0.00048 1
0.00054 1.06813308431902
0.0006 1.09268866293145
0.00066 1.11295566161721
0.00072 1.12962578681608
0.00078 1.14339074496784
0.00084 1.15476931590233
0.0009 1.16421110880542
0.00096 1.17213114754098
0.00102 1.17877152936294
0.00108 1.1844089368472
0.00114 1.18921629660372
0.0012 1.19336653524244
0.00126 1.19699799405132
0.00132 1.20017984367434
0.00138 1.20301584007747
0.00144 1.20557515390468
0.0015 1.20785778515598
0.00156 1.21000207511932
0.00162 1.21193885315072
0.00168 1.21377187521616
0.00174 1.21546655599364
0.0018 1.21709206612714
0.00186 1.21861382029467
0.00192 1.22010098914021
0.00198 1.22151898734177
0.00204 1.22290240022135
0.0021 1.22425122777893
0.00216 1.22556547001453
0.00222 1.22684512692813
0.00228 1.22809019851975
0.00234 1.22933527011136
0.0024 1.23054575638099
0.00246 1.23175624265062
0.00252 1.23296672892025
0.00258 1.23414262986788
0.00264 1.23531853081552
0.0027 1.23645984644117
0.00276 1.23760116206682
0.00282 1.23877706301446
0.00288 1.23988379331812
0.00294 1.24102510894376
};
\addlegendentry{Q1STc+}
\addplot [ultra thick, darkcyan0152161,dash pattern=on 4.4pt off 3.3pt]
table {%
0 0
6e-05 0.125198865601439
0.00012 0.250397731202878
0.00018 0.37698000968389
0.00024 0.501487168845542
0.0003 0.625994328007194
0.00036 0.750501487168846
0.00042 0.875008646330497
0.00048 1.00297433769108
0.00054 1.06868644947084
0.0006 1.09289617486339
0.00066 1.113647368057
0.00072 1.13094002905167
0.00078 1.14477415784741
0.00084 1.15514975444421
0.0009 1.16552535104102
0.00096 1.17244241543889
0.00102 1.17935947983676
0.00108 1.18627654423463
0.00114 1.18973507643356
0.0012 1.1931936086325
0.00126 1.19665214083143
0.00132 1.20011067303037
0.00138 1.2035692052293
0.00144 1.20702773742824
0.0015 1.20702773742824
0.00156 1.21048626962717
0.00162 1.21048626962717
0.00168 1.2139448018261
0.00174 1.2139448018261
0.0018 1.21740333402504
0.00186 1.21740333402504
0.00192 1.22086186622397
0.00198 1.22086186622397
0.00204 1.22432039842291
0.0021 1.22432039842291
0.00216 1.22432039842291
0.00222 1.22777893062184
0.00228 1.22777893062184
0.00234 1.22777893062184
0.0024 1.23123746282078
0.00246 1.23123746282078
0.00252 1.23469599501971
0.00258 1.23469599501971
0.00264 1.23469599501971
0.0027 1.23815452721865
0.00276 1.23815452721865
0.00282 1.23815452721865
0.00288 1.24161305941758
0.00294 1.24161305941758
};
\addlegendentry{U-P-SBFEM~\cite{sauren_mesh_2024}}
\addplot [ultra thick, darkslategray76, dash pattern=on 2.4pt off 3.2pt]
table {%
0 0
6e-05 0.125105485232068
0.00012 0.250183302206544
0.00018 0.375216158262433
0.00024 0.500242097253925
0.0003 0.625233450923428
0.00036 0.750190219270941
0.00042 0.875112402296465
0.00048 1
0.00054 1.06813308431902
0.0006 1.09268866293145
0.00066 1.11295566161721
0.00072 1.12962578681608
0.00078 1.14339074496784
0.00084 1.15476931590233
0.0009 1.16421110880542
0.00096 1.17213114754098
0.00102 1.17877152936294
0.00108 1.1844089368472
0.00114 1.18921629660372
0.0012 1.19336653524244
0.00126 1.19699799405132
0.00132 1.20017984367434
0.00138 1.20301584007747
0.00144 1.20557515390468
0.0015 1.20785778515598
0.00156 1.21000207511932
0.00162 1.21193885315072
0.00168 1.21377187521616
0.00174 1.21546655599364
0.0018 1.21709206612714
0.00186 1.21861382029467
0.00192 1.22010098914021
0.00198 1.22151898734177
0.00204 1.22290240022135
0.0021 1.22425122777893
0.00216 1.22556547001453
0.00222 1.22684512692813
0.00228 1.22809019851975
0.00234 1.22933527011136
0.0024 1.23054575638099
0.00246 1.23175624265062
0.00252 1.23296672892025
0.00258 1.23414262986788
0.00264 1.23531853081552
0.0027 1.23645984644117
0.00276 1.23760116206682
0.00282 1.23877706301446
0.00288 1.23988379331812
0.00294 1.24102510894376
};
\addlegendentry{Q1}
\end{axis}

\end{tikzpicture}
    \caption{Membrane patch test - Force-displacement curves for an elasto-plastic material, see Table~\ref{mat_param_plast}. Comparison between Q1STc, Q1STc+, the element U-P-SBFEM~\cite{sauren_mesh_2024} and the conventional low-order formulation using full integration (Q1). Here, the results of the proposed Q1STc+ element coincide with those of U-P-SBFEM and Q1.}
    \label{membrane_plasticity}
\end{figure}
\par
\begin{figure}[H]
	\centering
	\input{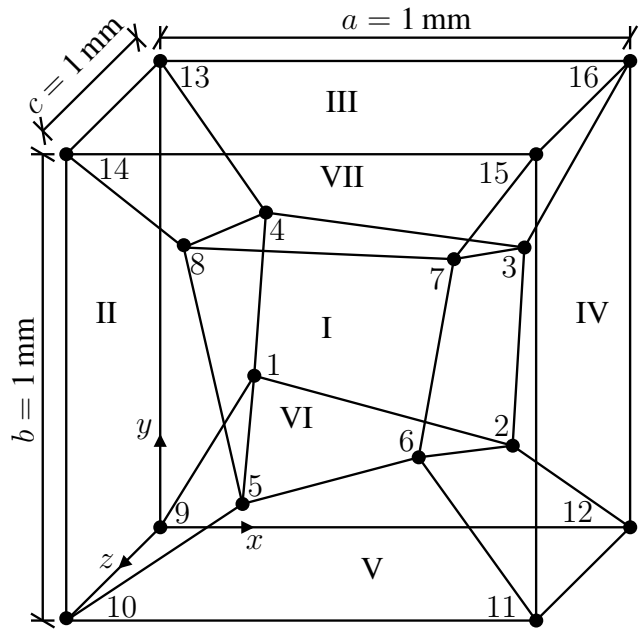}
	\caption{Patch test for solids - Geometry of the seven-element patch test with element numbers I - VII.}
	\label{Solid}
\end{figure}
\begin{figure}[H]
    \centering
    \def\svgwidth{0.7\textwidth}
    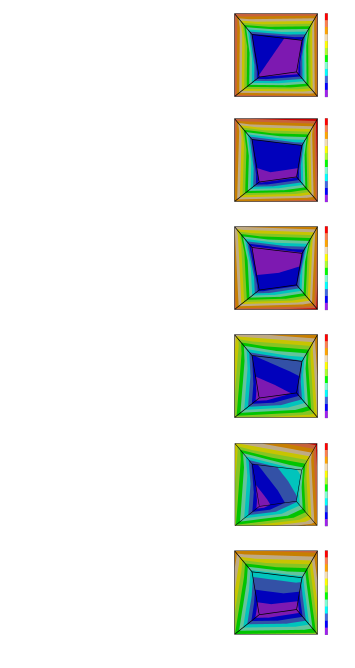
    \caption{Patch test for solids - Stress distribution at $z = \text{0.5}\,\text{mm}$ for Q1, Q1STc and Q1STc+, where Q1 denotes the low-order finite element formulation using full integration, and serves as an additional reference solution to the analytical solution in Table~\ref{tab_solid}.}
    \label{solid_stresses}
\end{figure}

\textbf{Patch test for solids.} The patch test for solids is depicted in Figure~\ref{Solid}. Here, a three dimensional $\text{1} \times \text{1} \times\text{1} \,[\text{mm}^3]$ solid is assumed, where the following displacements (in $\text{mm}$) at the edge nodes of the patch are prescribed as follows
\begin{equation}
    \bm{u} = \left(\begin{array}{c} u_x \\ u_y \\u_z\end{array}\right) = \left(\begin{array}{c} 0.001\,(2x+y+z)/2 \\ 0.001\,(x+2y+z)/2 \\ 0.001\,(x+y+2z)/2 \end{array}\right).
\end{equation}
For a St.Venant-Kirchhoff material behavior, the analytical solution of the normal and shear stresses in all three directions yield $\text{2001.5}\, \text{N}/\text{mm}^2$ and $\text{400.5}\, \text{N}/\text{mm}^2$, respectively. Table~\ref{tab_solid} shows the results obtained from the two element formulations. Figure~\ref{solid_stresses} shows the stress distribution for each element in the patch for both Q1STc and Q1STc+, where Q1 is used as a reference solution. The material parameters are $\lambda = \text{400000}$ $\text{N}/\text{mm}^2$ and $\mu = \text{400000} \, \text{N}/\text{mm}^2$ $(E = \text{10}^\text{6} \,\text{N}/\text{mm}^2, \nu = \text{0.25})$. It can be seen that both element formulations do not fulfill the patch test since the analytical solution is not obtained and furthermore, the stress is not homogeneous within the structure. In general, the results of Q1STc and Q1Stc+ behave similarly, since the error for the Q1STc element lies between $\text{0.03}\,\text{\%}- \text{15.8}\,\text{\%}$, while the error in the Q1STc+ element varies between $\text{0.04}\,\text{\%}$ and $\text{14.5}\,\text{\%}$. One possible reason why the membrane patch test is satisfied for Q1STc+, see Table~\ref{membrane_table}, while the patch test for solids is not, could be attributed to the fact that the membrane patch test is conducted on very thin structures $t = \text{0.001}\, \text{mm}$, where a plane stress state is assumed. In this case, the out-of-plane stresses are zero, and the behavior is dominated by in-plane stress and strain components. As a result, the influence of any out-of-plane effects in the $z$-direction is negligible. This simplification reduces the complexity of the test and makes it less sensitive to certain approximations in the element formulation. Consequently, the membrane patch test is more readily satisfied. A potential solution to address this could involve an extension of the Taylor series expansion to include the trilinear terms for the approximation of the strain and constitutive quantities. This may provide a more accurate representation and could lead to correct analytical results for the patch test for solids. 

\begin{table}[H]
	\centering
	\caption{Patch test for solids - Stress comparison between the analytical solution and results obtained using Q1STc and Q1STc+, see Figure~\ref{Solid}.} 
	\begin{tabular}{ c c c c c}
	\hline
        $\text{Stress [$\text{N}/\text{mm}^2$]}$ &	$\text{Element number}$	 & $\text{Analytical solution}$ & $\text{Q1STc}$ & $\text{Q1STc+}$ \\
	\hline
	\hline
        \multirow{7}{2em}{$S_{xx}$}	&	I	&	2001.50&	1715.20 & 1713.50\\
        &	II	&	2001.50&	2127.00 & 2149.60\\
        &	III	&	2001.50&	1969.50 & 1973.40\\
        &	IV	&	2001.50&	2167.50 & 2180.40\\
        &	V	&	2001.50&	1982.20 & 1961.20 \\
        &	VI	&	2001.50&	1971.70 & 1963.10\\
        &	VII	&	2001.50&	1972.20 & 1969.90\\
    \hline
        \multirow{7}{2em}{$S_{yy}$}	&	I	&	2001.50&	1687.40 & 1710.00\\
        &	II	&	2001.50&	1967.80 & 1989.90\\
        &	III	&	2001.50&	2090.60 & 2098.10\\
        &	IV	&	2001.50&	2000.80 & 2002.50\\
        &	V	&	2001.50&	2135.60 & 2092.30\\
        &	VI	&	2001.50&	1983.20 & 1980.40\\
        &	VII	&	2001.50&	1955.90 & 1964.00\\
    \hline
    \multirow{7}{2em}{$S_{zz}$}	&	I	&	2001.50&	1689.60 &1707.00\\
    &	II	&	2001.50&	1974.10 & 1988.70\\
    &	III	&	2001.50&	1963.20 & 1961.70\\
    &	IV	&	2001.50&	1993.80 & 2004.10\\
    &	V	&	2001.50&	1985.50 & 1982.60\\
    &	VI	&	2001.50&	2111.40 & 2084.30\\
    &	VII	&	2001.50&	2103.20 & 2114.30\\
	\hline	
    \multirow{7}{2em}{$S_{xy}$}	&	I	&	400.50&	403.96 &375.12\\
    &	II	&	400.50&	416.39  & 411.63 \\
    &	III	&	400.50&	401.03 & 405.65\\
    &	IV	&	400.50&	396.76 & 419.72\\
    &	V	&	400.50&	389.54& 401.84\\
    &	VI	&	400.50&	391.83 & 389.29\\
    &	VII	&	400.50&	407.83 & 393.07\\
	\hline
    \multirow{7}{2em}{$S_{xz}$}	&	I	&	400.50&	378.29 &367.89\\
    &	II	&	400.50&	375.52 & 388.84\\
    &	III	&	400.50& 401.58& 395.17\\
    &	IV	&	400.50&	400.07 & 399.39\\
    &	V	&	400.50&	388.65& 389.43\\
    &	VI	&	400.50&	420.10 & 419.18\\
    &	VII	&	400.50&	426.18 & 427.03\\
	\hline
    \multirow{7}{2em}{$S_{yz}$}	&	I	&	400.50&	350.34 &362.46\\
    &	II	&	400.50&	386.41 & 391.73\\
    &	III	&	400.50&	414.28& 409.72\\
    &	IV	&	400.50&	384.76 & 390.46\\
    &	V	&	400.50&	386.64& 396.61\\
    &	VI	&	400.50&	411.80 & 409.13\\
    &	VII	&	400.50&	438.12 & 421.17\\
	\hline
	\end{tabular}
    \label{tab_solid}
	\end{table}

\subsection{Cube under compression}
\label{cube_under_comp_ex}
\begin{figure}[H]
    \centering
    \def\svgwidth{0.5\textwidth}
\begingroup%
  \makeatletter%
  \providecommand\color[2][]{%
    \errmessage{(Inkscape) Color is used for the text in Inkscape, but the package 'color.sty' is not loaded}%
    \renewcommand\color[2][]{}%
  }%
  \providecommand\transparent[1]{%
    \errmessage{(Inkscape) Transparency is used (non-zero) for the text in Inkscape, but the package 'transparent.sty' is not loaded}%
    \renewcommand\transparent[1]{}%
  }%
  \providecommand\rotatebox[2]{#2}%
  \newcommand*\fsize{\dimexpr\f@size pt\relax}%
  \newcommand*\lineheight[1]{\fontsize{\fsize}{#1\fsize}\selectfont}%
  \ifx\svgwidth\undefined%
    \setlength{\unitlength}{198.42519685bp}%
    \ifx\svgscale\undefined%
      \relax%
    \else%
      \setlength{\unitlength}{\unitlength * \real{\svgscale}}%
    \fi%
  \else%
    \setlength{\unitlength}{\svgwidth}%
  \fi%
  \global\let\svgwidth\undefined%
  \global\let\svgscale\undefined%
  \makeatother%
  \begin{picture}(1,1.07142857)%
    \lineheight{1}%
    \setlength\tabcolsep{0pt}%
    \put(0,0){\includegraphics[width=\unitlength,page=1]{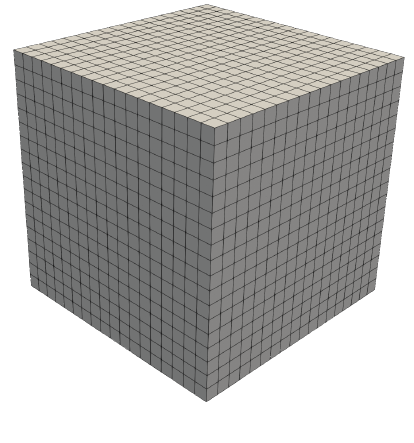}}%
    \put(0.6049057,-0.00517556){\makebox(0,0)[lt]{\lineheight{1.25}\smash{\begin{tabular}[t]{l}$x$\end{tabular}}}}%
    \put(0,0){\includegraphics[width=\unitlength,page=2]{cube_imag.pdf}}%
    \put(0.47789961,0.24724764){\makebox(0,0)[lt]{\lineheight{1.25}\smash{\begin{tabular}[t]{l}$y$\end{tabular}}}}%
    \put(0.34978995,-0.00517556){\makebox(0,0)[lt]{\lineheight{1.25}\smash{\begin{tabular}[t]{l}$z$\end{tabular}}}}%
    \put(0,0){\includegraphics[width=\unitlength,page=3]{cube_imag.pdf}}%
    \put(0.4892098,0.81079764){\makebox(0,0)[lt]{\lineheight{1.25}\smash{\begin{tabular}[t]{l}$\textcolor{white}{u_y}$\end{tabular}}}}%
  \end{picture}%
\endgroup%

    \caption{Cube under compression - Geometry and boundary value problem with a discretization of 18x18x18 = 5832  $\mathrm{n_{el}}$ elements. The surface load $p_0 = \text{100}\,\text{N}/\text{mm}^2$ is applied on one-quarter of the top block indicated by the dark gray area. The red cross marks the displacement $u_y$ which will be used as a measure to investigate the results obtained for the convergence study and contour plots. Due to symmetry, only one quarter of the structure is modeled, where symmetry conditions have to be considered for the planes $x = \text{0}$ and $z = \text{0}$.}
    \label{cube_imag}
\end{figure}

While Q1 satisfies both the membrane patch test, see Figure~\ref{membrane_stresses}, and the patch test for solids, see Figure~\ref{solid_stresses}, ensuring it can reproduce a homogeneous stress state, this alone does not guarantee the absence of issues such as volumetric or shear locking. Passing the patch test primarily ensures that the element is able to produce a homogeneous stress state, see~\cite{macneal_proposed_1985}. However, locking usually occurs for complex examples. To further test the performance of Q1STc+ and its benefits in comparison to the standard finite element formulation using full integration (Q1), a typical example for volumetric locking is shown. For this, a cube with side lengths of $\text{1} \,\text{mm}$ is subjected to compression, see Figure~\ref{cube_imag}. Since only one quarter of the system is discretized, the symmetry conditions have to be considered on the planes $x = \text{0}$ and $z = \text{0}$. A surface load of $p_0 = \text{100}\,\text{N}/\text{mm}^2$ is applied linearly in the dark gray area on the top of the block. The nodes on the top of the structure are constrained in $x$- and $z$- direction. The chosen material model is a compressible Neo-Hooke model with the following Helmholtz free energy
\begin{equation}
    \psi = \frac{\mu}{2}\,(\trace{(\bm{C})} - 3 - \ln{(\det{(\bm{C})})}) + \frac{\lambda}{4} \, (\det{(\bm{C})} - 1 -\ln{(\det{(\bm{C})})}).
\end{equation}
Here, the material parameters are chosen as $\mu =\text{80.194}\,\text{N}/\text{mm}^2$ and $\lambda = \text{40016.806}\,\text{N}/\text{mm}^2$.
To investigate how Q1 and Q1STc+ perform for an example typically known to challenge issues of volumetric locking, a convergence study for a regular mesh was conducted, see Figure~\ref{cube_comp}. Here, the simulation was performed for a discretization of 512, 1000, 1728 and 5832 $\mathrm{n_{el}}$ elements, where the elements in each coordinate direction were evenly distributed. The results of the absolute displacement $\lvert u_y\rvert$ at the node indicated by the red cross in Figure~\ref{cube_imag} at the end of the simulation were plotted. An indication for locking in general is an underestimation of the displacements which at the end results in a stiffer response of the structure and influences the convergence behavior. Q1 proves this by showing a typical locking behavior in Figure~\ref{cube_comp}, where no convergence is reached even for the finest discretization of 18x18x18 = 5832 $\mathrm{n_{el}}$ elements. In contrast, for Q1STc+, convergence can be observed even for a mesh of 10x10x10 = 1000 $\mathrm{n_{el}}$ elements. Figure~\ref{cube_paraview} shows the contours of the displacement $u_y$ of the cube at the end of the simulation. Here, the results are shown for a discretization of 18x18x18 = 5832 $\mathrm{n_{el}}$ elements. As concluded in Figure~\ref{cube_comp}, a clear underestimation of the displacements can be observed for Q1. For this reason, the results of a volumetric locking-free element formulation are used as an additional comparison for the next structural example, which will be further discussed in detail in the next Section.
\begin{figure}[H]
    \centering
\begin{tikzpicture}

\definecolor{darkgray176}{RGB}{176,176,176}
\definecolor{darkslategray76}{RGB}{76,76,76}
\definecolor{lightgray204}{RGB}{204,204,204}
\definecolor{orange2461680}{RGB}{246,168,0}

\begin{axis}[
legend cell align={left},
legend style={
  fill opacity=1,
  draw opacity=1,
  text opacity=1,
  at={(0.97,0.03)},
  anchor=south east,
  draw=lightgray204,
  font=\fontsize{11}{9}\selectfont,
},
legend image post style={mark=square*},
tick align=outside,
tick pos=left,
x grid style={darkgray176},
xlabel={Number of elements $\mathrm{n_{el}}$},
xmajorgrids,
xmin=246, xmax=6098,
xtick style={color=black},
y grid style={darkgray176},
ylabel={Displacement \(\lvert\displaystyle u_y\rvert\) [$\text{mm}$]},
ymajorgrids,
ymin=0.2837525, ymax=0.5228375,
ytick style={color=black}
]
\addplot [ultra thick, darkslategray76,mark=square*, mark size=2,mark options={solid}]
table {%
512 0.29462
1000 0.34098
1728 0.37672
5832 0.44312
};
\addlegendentry{Q1}
\addplot [ultra thick, orange2461680,mark=square*, mark size=2,mark options={solid}]
table {%
512 0.50823
1000 0.51078
1728 0.51097
5832 0.51197
};
\addlegendentry{Q1STc+}
\end{axis}

\end{tikzpicture}
    \caption{Cube under compression - Convergence study of the conventional low-order formulation using full integration Q1 and Q1STc+. Here, the absolute value of $u_y$ of the node marked by the red cross in Figure~\ref{cube_imag} is used as a measure and plotted over 512, 1000, 1728 and 5832 $\mathrm{n_{el}}$ elements at the end of the simulation. The elements are evenly distributed over the coordinate directions.}
    \label{cube_comp}
\end{figure}
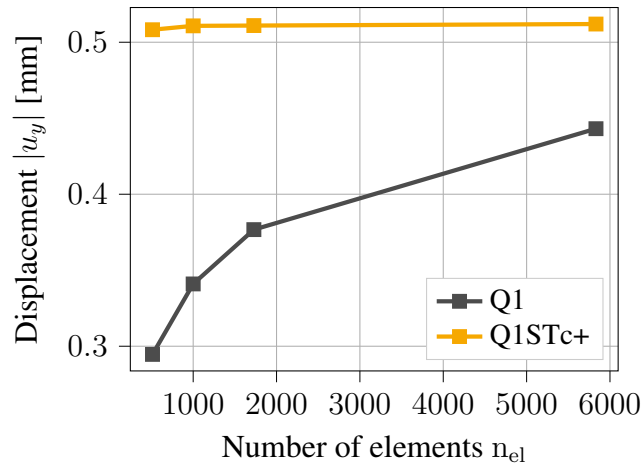
\begin{figure}[H]
    \centering
    \def\svgwidth{0.9\textwidth}
    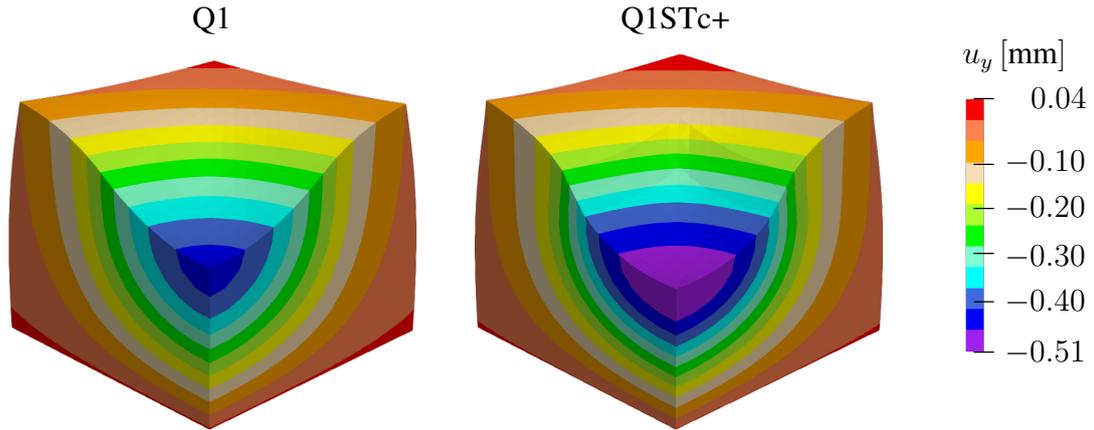
    \caption{Cube under compression - Contours of the displacement $u_y$ for a discretization of 18x18x18 = 5832 $\mathrm{n_{el}}$ elements at the end of the simulation. Here, the perspective aligns with the coordinate system depicted in Figure~\ref{cube_imag}.}
    \label{cube_paraview}
\end{figure}

\subsection{Asymmetrically notched specimen}
\begin{figure}[H]
	\centering
	\input{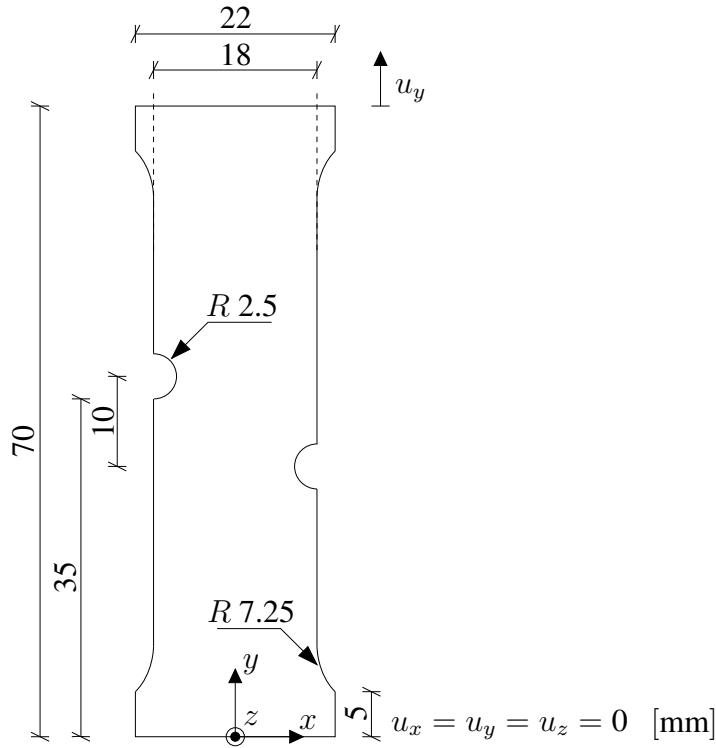}
	\caption{Asymmetrically notched specimen - Geometry and boundary value problem, with $z = \text{3} \,\text{mm}$.}
	\label{ANS}
\end{figure}

It is now investigated, whether the performance of Q1STc and Q1STc+ differs for a more common structural example, particularly under the influence of plastic material behavior. For this, an asymmetrically notched specimen is clamped at the bottom part ($u_x = u_y = u_z=\text{0}\, \text{mm}$) and subjected to a vertical displacement at the top part, see~\cite{ambati_phase-field_2016}. The geometry and the boundary value problem are shown in Figure~\ref{ANS}, while the chosen material parameters are illustrated in Table~\ref{mat_param_plast_ANN}. The total length of the specimen is $\text{70}\, \text{mm}$ with a thickness of $\text{3} \,\text{mm}$.\par
The Q1STc element formulation did not pass the membrane patch test, see Section~\ref{patch_tests}. In order to see whether a distortion of elements influences the results for Q1STc, the elements of the specimen were slightly distorted without altering the overall geometry of the structure. The elements were randomly distorted in $x$- and $y$-direction. Two degrees of distortion are considered: a random distortion between $\mathrm{d} = -\text{0.2}\,\text{mm}$ and $\mathrm{d} = \text{0.2}\,\text{mm}$, and a random distortion between $\mathrm{d} = -\text{0.5}\,\text{mm}$ and $\mathrm{d} = \text{0.5}\,\text{mm}$. \par Neither Q1STc nor Q1STc+ satisfied the patch test for solids, see Section~\ref{patch_tests}. To investigate whether this limitation could influence the performance on a structural level, including in the $z$-direction, the same procedure for introducing distortions was applied in the $z$-direction, denoted as $\mathrm{d_z} = \pm\,\text{0.2}\, \text{mm}$ and $\mathrm{d_z} = \pm\,\text{0.5}\,\text{mm}$. The purpose of this example serves to demonstrate the performance of the two element formulations Q1STc and Q1STc+, in particular considering the case where the mesh is distorted. \par
\begin{figure}[H]
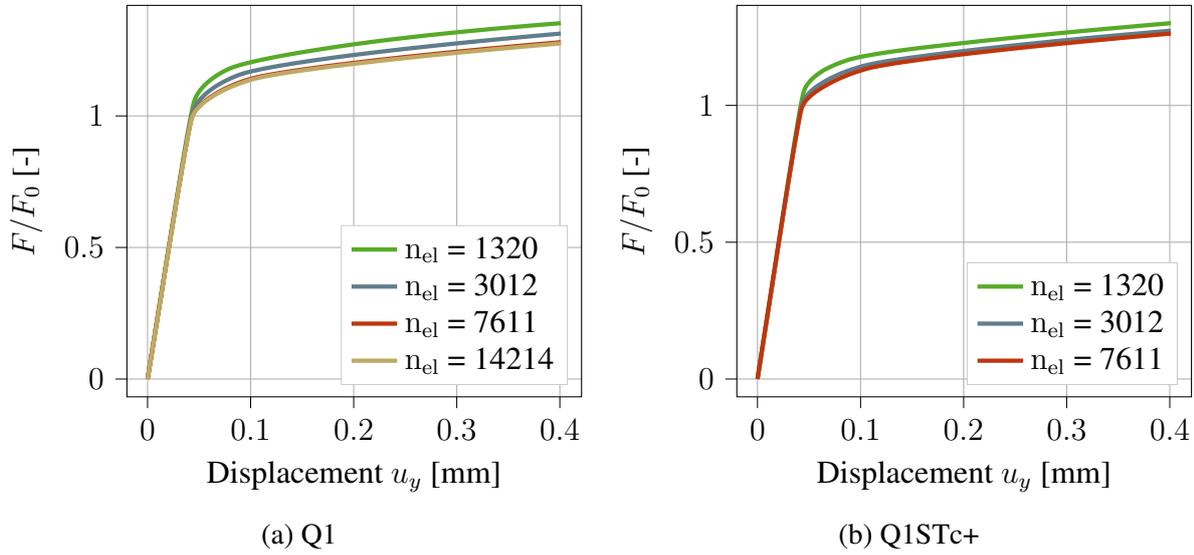

    	\centering
    	\begin{subfigure}[t]{0.495\textwidth}
            \centering
    		\input{data/ANS/Konvergenz/Konvergenz_Q1.tex}
    		\caption{Q1}
    		\label{Conv_Q1}
    	\end{subfigure}
    	\begin{subfigure}[t]{0.495\textwidth}
            \centering
            \input{data/ANS/Konvergenz/Konvergenz_Q1STc+.tex}
    		\caption{Q1STc+}
    		\label{Conv_Q1STc+}
    	\end{subfigure}
    \caption{Asymmetrically notched specimen - Convergence studies on the standard Q1 (a) and the Q1STc+ (b) formulation for an elasto-plastic material.}
    \label{Convergence}
    \end{figure}

 As was done for the patch tests, see Section~\ref{patch_tests}, the performance of the standard low-order finite element formulation (Q1) using eight Gauss points is also investigated on this structural example, since the derivation of both element formulations Q1STc and Q1STc+ are based on it. Before comparing the results of Q1STc and Q1STc+, convergence studies on various meshes, where three elements were used to model the thickness direction, were done with Q1 and Q1STc+, see Figure~\ref{Convergence}. For a simpler illustration, all curves in this section are normalized by $F_0 = \text{3972.8}\,\text{N}$, where $F_0$ denotes the largest force obtained by the Q1STc+ formulation for $\mathrm{d} = \mathrm{d_z} = \text{0}\,\text{mm}$ in the elastic regime. Several convergence studies for different structural examples, including the asymmetrically notched specimen, were also carried out by~\citet{barfusz_single_2021} for the Q1STc formulation. For both element formulations, convergence can be observed by increasing the mesh density. The force-displacement curves for different mesh densities show much closer agreement with the converged solution when using Q1STc+ compared to the standard Q1 formulation. A possible explanation for this is the restriction of the material model to volume-preserving plasticity, see Section~\ref{plasti_theory}. Volume-preserving plasticity assumes the material undergoes incompressible plastic deformation, which imposes a constraint that can amplify the likelihood of volumetric locking. This issue is particularly pronounced in conventional low-order finite element formulations with full integration, as it often leads to artificially stiff element behavior and influences the convergence. Additionally, the asymmetrically notched specimen introduces further challenges due to the complex interplay of tensile and shear stresses near the notch tip, which could also influence the convergence behavior. For the standard Q1 formulation, the mesh with 14214 elements will be used for comparison for the following simulations since convergence is not achieved with 3012 elements.
\begin{table}[H]
    \centering
    \caption{Asymmetrically notched specimen -  Material parameters for an elasto-plastic material, see Figure~\ref{ANS}.}
\begin{tabular}{ l l l l}
\hline
                    $\text{Symbol}$ &	$\text{Material parameter}$	 & $\text{Value}$& $\text{Unit}$	\\
\hline
\hline
    $\lambda$	&	 First Lam\'{e} parameter & 25000.0	& $\text{N}/\text{mm}^2$\\
    $\mu$	&	 Second Lam\'{e} parameter& 55000.0	& $\text{N}/\text{mm}^2$\\
    $a$	&	 First kinematic hardening stiffness parameter & 62.5	&$\text{N}/\text{mm}^2$\\
    $b$	&	 Second kinematic hardening stiffness parameter&2.5	& -\\
    $e$	&	 First isotropic hardening parameter &125.0	&$\text{N}/\text{mm}^2$\\
    $f$	&	 Second isotropic hardening parameter & 5.0	&-\\
    $\sigma_{y0}$	&	 Initial plastic threshold & 100.0	&$\text{N}/\text{mm}^2$\\
\hline	
\label{mat_param_plast_ANN}
\end{tabular}
\end{table}

\begin{figure}[H]
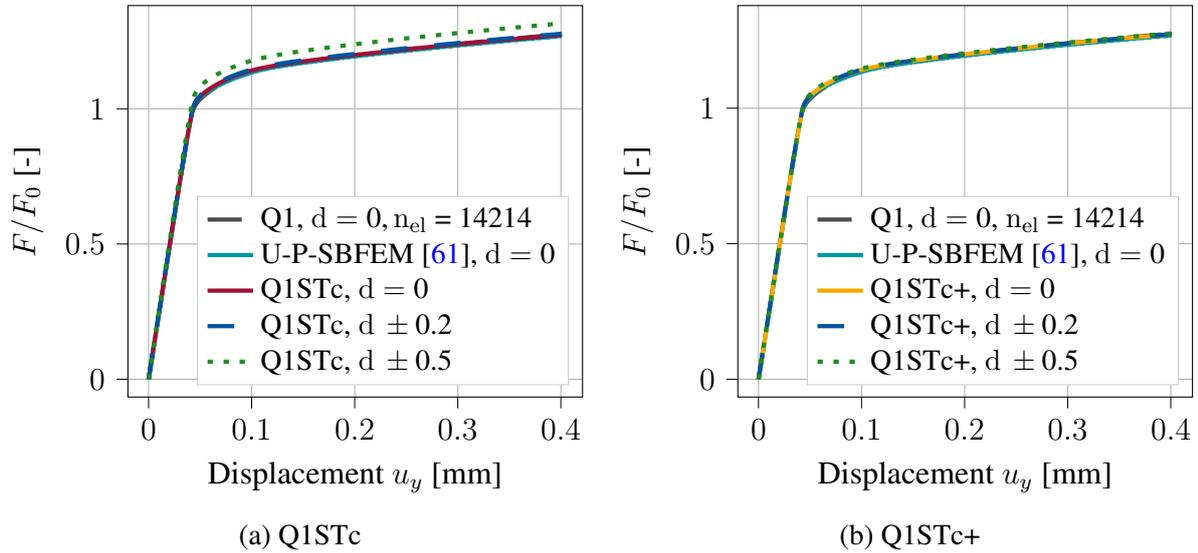

    	\centering
    	\begin{subfigure}[t]{0.495\textwidth}
            \centering
    		\input{data/ANS/3012/xy/3012_Q1Stc_xy_split_new.tex}
    		\caption{Q1STc}
    		\label{Q1STc_xy}
    	\end{subfigure}
    	\begin{subfigure}[t]{0.495\textwidth}
            \centering
            \input{data/ANS/3012/xy/3012_Q1Stc+_xy_split_new.tex}
    		\caption{Q1STc+}
    		\label{Q1STc+_xy}
    	\end{subfigure}
    \caption{Asymmetrically notched specimen - Force-displacement curves of Q1STc (a) and Q1STc+ (b) for an elasto-plastic material, with a distortion in x- and y-direction and a mesh consisting of 3012 elements, while $\mathrm{d_z} =\text{0}\,\text{mm}$. The finite element formulation (Q1) and the mixed finite element formulation based on scaled boundary parametrization (U-P-SBFEM) with a mesh consisting of 14214 elements and 3012 elements are used for comparison, respectively.}
    \label{3012_distortion_xy}
    \end{figure}

Figure~\ref{3012_distortion_xy} shows the performance of Q1STc and the Q1STc+ for a mesh distortion in x- and y-direction. For more common benchmark examples which are designed to study the locking phenomena, e.g. "Cook's membrane" or "Cube under compression", the standard Q1 element formulation has the tendency to show locking behavior. This was proved in Section~\ref{cube_under_comp_ex} where Q1 showed an underestimation of the displacements, resulting in a much stiffer structural response, see Figures~\ref{cube_comp} and~\ref{cube_paraview}. Since the underlying example is combined with volume-preserving plasticity and thus presents volumetric-locking in the plastic regime, see Figure~\ref{Conv_Q1}, the conventional Q1 cannot be used as a reliable reference to see whether the results of Q1STc and Q1STc+ are plausible. To this end, as was done for the membrane patch test above, the volumetric locking-free mixed scaled boundary finite element formulation of \citet{sauren_mixed_2023, sauren_stability_2024,sauren_mesh_2024}, extended with our material formulation provided in Section~\ref{plasti_theory}, with 3012 elements was used for comparison. For Q1STc, the distortion of the mesh has an influence on the results with $\mathrm{d}\,\pm\text{0.5}\,\text{mm}$, see Figure~\ref{Q1STc_xy}, which in general should not be the case. In comparison, this is not the case for Q1STc+, see Figure~\ref{Q1STc+_xy}. Both Q1STc and Q1STc+ show plausible results when distorting the elements only in the thickness direction, see Figure~\ref{3012_distortion_z}. Figure~\ref{3012_distortion_xyz} shows the force-displacement curves of Q1STc and Q1STc+, where the mesh was distorted in all three directions. Here, the curves of Q1 and U-P-SBFEM are used for comparison. Although a shifting in only z-direction has no influence on the performance of Q1STc, it does contribute towards different results when also taking the other two directions into consideration compared to the results presented in Figure~\ref{3012_distortion_xy}. Figures~\ref{ANS_d_0_para} and~\ref{ANS_d_2_para} show the evolution of the accumulated plastic strain $\kappa$ at three different loading stages, which are marked in Figures~\ref{Q1STc_xyz} and~\ref{Q1STc+_xyz}, where the latter shows the results for a distorted mesh with $\mathrm{d}\,\pm\text{0.2}\,\text{mm}$ and $\mathrm{d_z}\,\pm\text{0.5}\,\text{mm}$. Overall it can be seen that there is no significant difference between the results obtained without a distorted mesh and with a distorted one with Q1STc and Q1STc+, as was the case when investigating the force-displacement curves, see Figure~\ref{3012_distortion_xyz}.

    \begin{figure}[H]
    	\centering
    	\begin{subfigure}[t]{0.495\textwidth}
            \centering
    		\input{data/ANS/3012/z/3012_Q1Stc_split_z_new.tex}
    		\caption{Q1STc}
    		\label{Q1STc_z}
    	\end{subfigure}
    	\begin{subfigure}[t]{0.495\textwidth}
            \centering
            \input{data/ANS/3012/z/3012_Q1Stc+_split_z_new.tex}
    		\caption{Q1STc+}
    		\label{Q1STc+_z}
    	\end{subfigure}
    \caption{Asymmetrically notched specimen - Force-displacement curves of Q1STc (a) and Q1STc+ (b) for an elasto-plastic material, with a distortion in z-direction and a mesh consisting of 3012 elements, while $\mathrm{d}= \text{0}\,\text{mm}$. The finite element formulation (Q1) and the mixed finite element formulation based on scaled boundary parametrization (U-P-SBFEM) with a mesh consisting of 14214 elements and 3012 elements are used for comparison, respectively.}
    \label{3012_distortion_z}
    \end{figure}
    
    \begin{figure}[H]
    	\centering
    	\begin{subfigure}[t]{0.495\textwidth}
            \centering
    		\input{data/ANS/3012/xyz/3012_Q1Stc_split_xyz_new.tex}
    		\caption{Q1STc}
    		\label{Q1STc_xyz}
    	\end{subfigure}
    	\begin{subfigure}[t]{0.495\textwidth}
            \centering
            \input{data/ANS/3012/xyz/3012_Q1Stc+_split_xyz_new.tex}
    		\caption{Q1STc+}
    		\label{Q1STc+_xyz}
    	\end{subfigure}
    \caption{Asymmetrically notched specimen - Force-displacement curves of Q1STc (a) and Q1STc+ (b) for an elasto-plastic material, with a distortion in x-, y- and z-direction and a mesh consisting of 3012 elements, while $\mathrm{d_z}\,\pm \text{0.5}\,\text{mm}$. The finite element formulation (Q1) and the mixed finite element formulation based on scaled boundary parametrization (U-P-SBFEM) with a mesh consisting of 14214 elements and 3012 elements are used for comparison, respectively. Here, $\mathrm{d_z}= \text{0}\,\text{mm}$ holds for Q1 and UP-SBFEM.}
    \label{3012_distortion_xyz}
    \end{figure}

\begin{figure}[H]
    \centering
    \def\svgwidth{0.9\textwidth}
    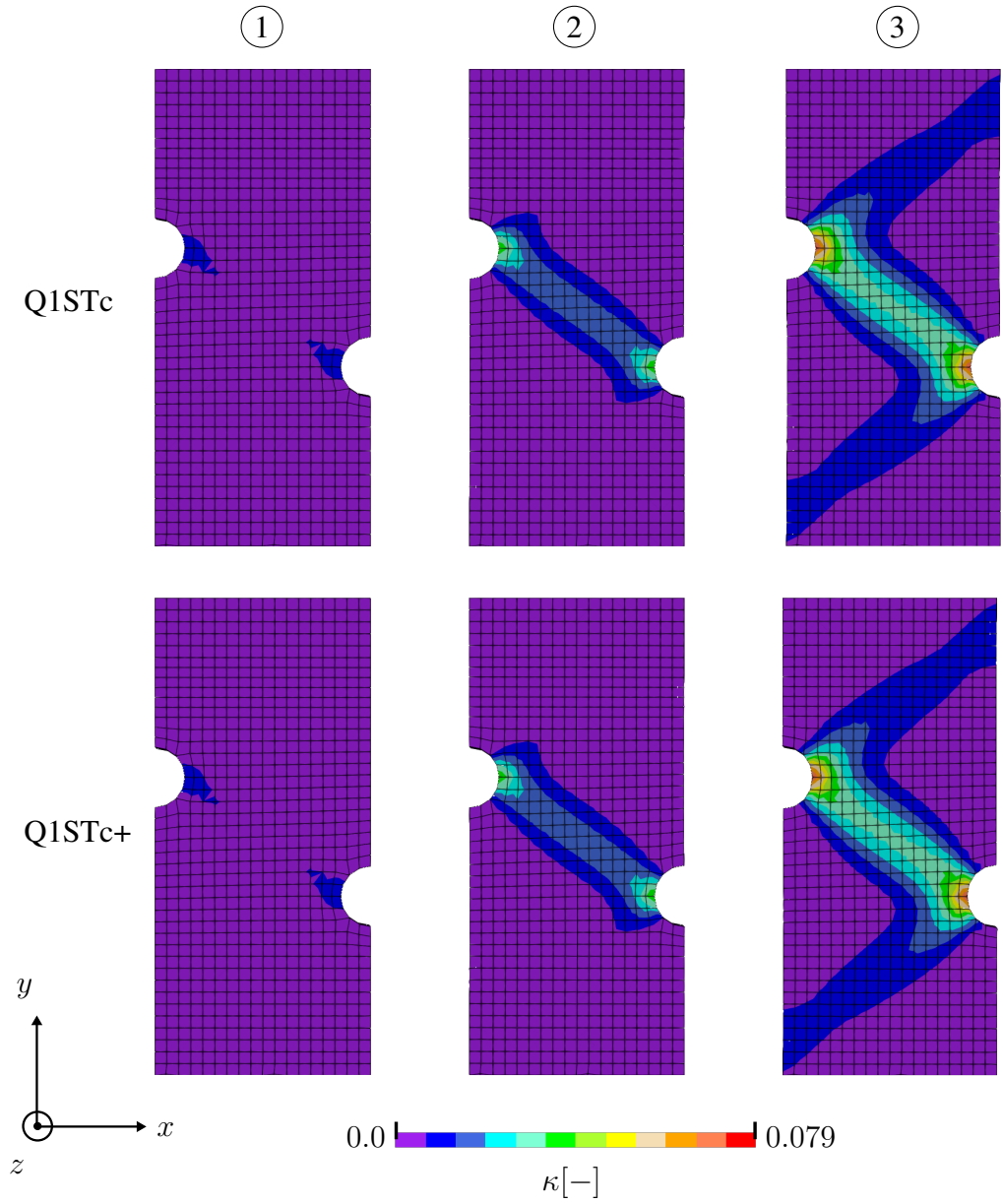
    \caption{Asymmetrically notched specimen - Contours of the accumulated plastic strain $\kappa$ at three different loading stages, see Figure~\ref{3012_distortion_xyz}, with 3012 elements without mesh distortion.}
    \label{ANS_d_0_para}
\end{figure}

\begin{figure}[H]
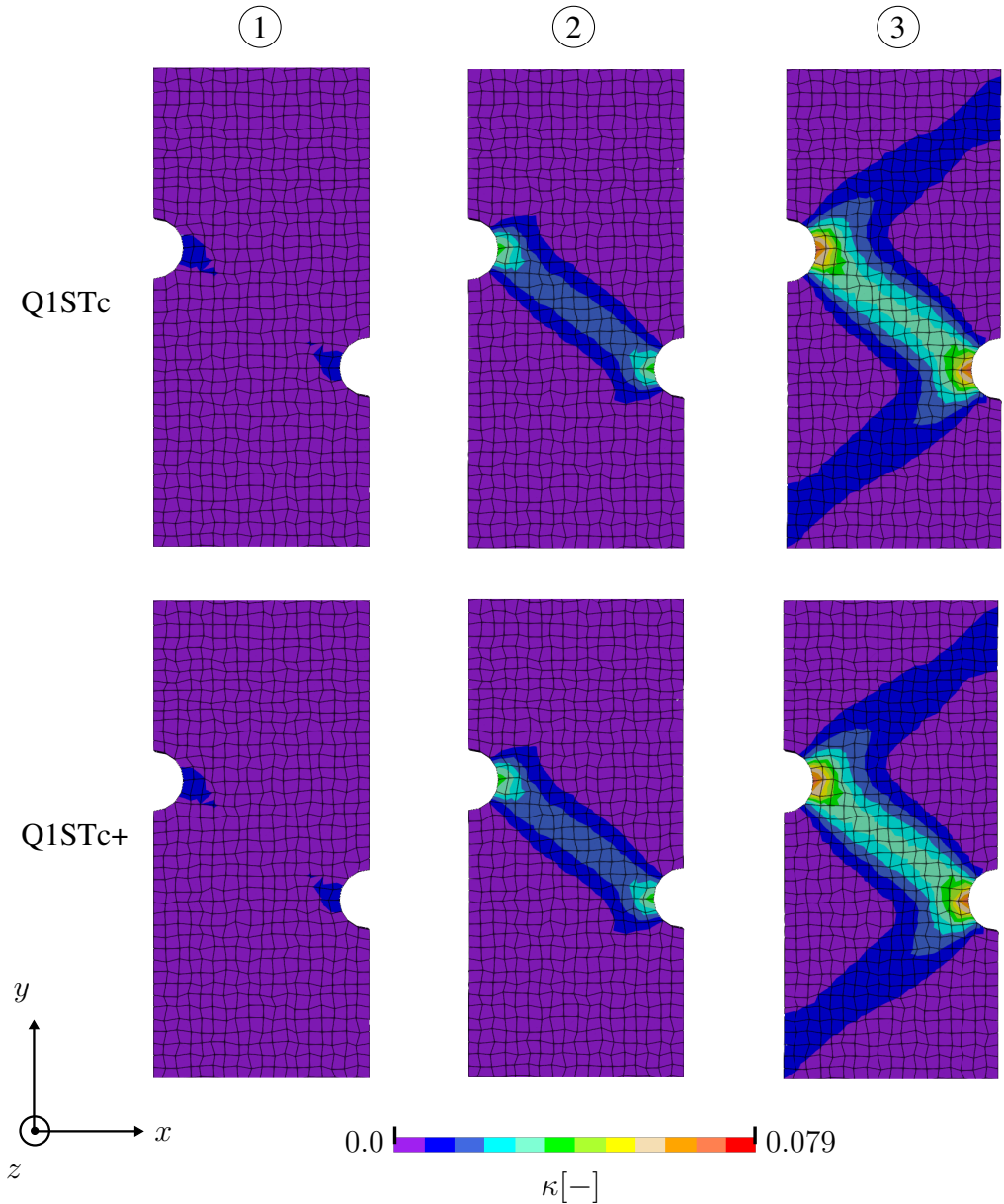

    \centering
    \def\svgwidth{0.9\textwidth}
    
    \caption{Asymmetrically notched specimen - Contours of the accumulated plastic strain $\kappa$ at three different loading stages, see Figure~\ref{3012_distortion_xyz}, with 3012 elements and mesh distortion of $\mathrm{d}\,\pm\text{0.2}\,\text{mm}$ and $\mathrm{d_z}\,\pm \text{0.5}\,\text{mm}$.}
    \label{ANS_d_2_para}
\end{figure}

\section{Conclusion}
\label{conclusion}
An enhanced single Gaussian point locking-free finite element formulation, termed Q1STc+, was presented, where the geometry of the system was calculated exactly using AD, instead of the need of an approximation by using a Taylor series expansion. This modification was carried out on the basis of the hypothesis that more accurate results would be obtained for certain tests, which proved to be true since this improvement leads to the fact that the membrane patch test for a St.Venant Kirchhoff material was fulfilled. The same simulation was also conducted with the difference that an elasto-plastic material was used. The results of Q1STc were not in accordance with Q1STc+ and U-P-SBFEM. To further investigate mesh distortion on a three dimensional case, the patch test for solids was applied. Both Q1STc and Q1STc+ did not fulfill the patch test. The reason for this could lie in the non-sufficient approximation of the strain and constitutive quantities by a Taylor series expansion. A possible solution to overcome this problem would be an expansion up to trilinear terms. While the conventional low-order finite element formulation using full integration with eight Gauss points (Q1) satisfies both patch tests, see Section~\ref{patch_tests}, ensuring it can reproduce a homogeneous stress state, this alone does not guarantee the absence of issues that can occur on a structural level. Typically, Q1 exhibits shear and volumetric locking. To investigate this phenomenon, a cube--a common example for studying volumetric locking--was subjected to compression, and a convergence study was performed on Q1STc+ and Q1, see Section~\ref{cube_under_comp_ex}. The results revealed typical locking behavior for Q1, where convergence was not achieved even for a very fine mesh. To further investigate the fact that the performance of Q1STc depends on the magnitude of the mesh distortion, an asymmetrically notched specimen was subjected to tension and its results were compared to the Q1STc+ element formulation. Differently distorted meshes were used to see the influence on the performance. The results showed that the force-displacement curves of Q1STc were not aligned with the results obtained from Q1, Q1STc+ and U-P-SBFEM, and exhibited variations depending on the severity of the mesh distortion. Overall, the enhanced Q1STc element formulation leads to significant improvements regarding mesh distortion. Since the patch test for solids was not fulfilled, this requires further investigation and improvement. In addition to that, the derivations in Section~\ref{plasti_theory}, in particular Equation~\eqref{Svol_press} and Appendix~\ref{SBFEM_theorie} were only proven for the case of volume-preserving plasticity, where $J_p=\text{1}, J_e = J$. Whether it also holds for non-volume-preserving plasticity models, such as the Drucker--Prager model~\cite{drucker_soil_1952}, will be subject of future works.

\addtocontents{toc}{\vspace{2em}} 

\appendix 
\numberwithin{equation}{section}


\section{Appendix A. Additions}
\label{appendix}
\subsection{Overview of the volumetric locking-free mixed displacement pressure formulation}
\label{SBFEM_theorie}
\textbf{Derivation of the true pressure.} In Section~\ref{plasti_theory}, see Equation~\eqref{Svol_press}, the true pressure is derived in detail as follows
\begin{align}
  \begin{split}
    p &= \frac{\partial \psi_e^{\mathrm{vol}}}{\partial J} = \frac{\partial \psi_e^{\mathrm{vol}}}{\partial J_e}\cdot \frac{\partial J_e}{\partial J_p} \\
    &= \kappa\, \Bigl(\sqrt{\det{(\bar{\bm{C}}}_e)} - 1\Bigr)\,\frac{1}{\underbrace{J_p}_{= 1}}\\
    &= \kappa\, \Bigl(\underbrace{\sqrt{\det{\bm{U}_p^{-1}\bm{C}\,\bm{U}_p^{-1}}}}_{= J} - 1\Bigr)\\
    &= \kappa\,(J-1).
  \end{split}
\end{align}\par
\textbf{Derivation of the volumetric part of the second Piola-Kirchhoff stress tensor.} The volumetric part of the second Piola-Kirchhoff stress tensor can be obtained as follows 
\begin{align}
  \begin{split}
  \bm{S}^{\mathrm{vol}} &= 2\,\bm{U}_p^{-1}\,\frac{\partial \psi_e^{\mathrm{vol}}}{\partial \bar{\bm{C}}_e}\,\bm{U}_p^{-1}\\
  &= \kappa\, \Bigl(\sqrt{\det{(\bar{\bm{C}}}_e)} - 1\Bigr)\,\sqrt{\det{(\bar{\bm{C}}}_e)}\,\bm{U}_p^{-1} \, \bar{\bm{C}}_e^{-1}\,\bm{U}_p^{-1}\\
  &= \kappa\, \Bigl(\underbrace{\sqrt{\det{\bm{U}_p^{-1}\bm{C}\,\bm{U}_p^{-1}}}}_{= J} - 1\Bigr)\, \underbrace{\sqrt{\det{\bm{U}_p^{-1}\bm{C}\,\bm{U}_p^{-1}}}}_{= J}\,\underbrace{\bm{U}_p^{-1} \,\bm{U}_p^{-1}}_{\bm{= I}}\bm{C}^{-1}\,\underbrace{\bm{U}_p^{-1}\,\bm{U}_p^{-1}}_{\bm{= I}}\\
  &= \underbrace{\kappa\,(J-1)}_{= p}\,J\,\bm{C}^{-1}\\
  &= p\,J\,\bm{C}^{-1}.
  \end{split}
  \label{A1_Svol}
\end{align}
It is important to note that these derivations hold for volume-preserving plasticity, meaning that $\sqrt{\det{\bar{\bm{C}}}_e} = J_e = J = \sqrt{\det{\bm{C}}} $, since $J_p = 1 $ and $J = J_e\,J_p$. \par
\textbf{Volumetric locking-free mixed element formulation -- Overview.} Since these formulations exhibit a nearly incompressible material behavior in the plastic regime, a volumetric locking-free solution is herein obtained using the formulations presented in \cite{sauren_mixed_2023, sauren_stability_2024, sauren_mesh_2024}. The formulation builds on the fact that the incompressibility constraint is enforced using a perturbed Lagrangian, leading to a linearized saddle-point problem of the form 
\begin{subequations}\label{P4E22}
	\begin{align}
		a(\delta\bm{u},\ \Delta \bm{u}) + b(\delta\bm{u},\ \Delta p) & =  
		r_u(\delta\bm{u}),  \\
		b(\delta p,\ \Delta \bm{u}) - c(\delta p,\ \Delta p) & =  
		r_p(\delta p) 
	\end{align}
\end{subequations}
with
\begin{subequations}\label{P4E15}
	\begin{align}
		a(\delta\bm{u},\ \Delta \bm{u}) & = 
		\int_{\Omega} \delta\bm{E} : \bm{\mathbb{C}} : \Delta\bm{E}\ \mathrm{d}V
		+ \int_{\Omega} \bm{S}: \Delta\delta\bm{E}\ \mathrm{d}V, \\ \label{P4E15b} 
		b(\delta\bm{u},\ \Delta p) & = \int_{\Omega} \delta\bm{E} : \bm{S}'_\mathrm{vol}\ \Delta p\ \mathrm{d}V, \\
		b(\delta p,\ \Delta \bm{u}) &= 
		\int_{\Omega} \delta p \ \bm{S}'_\mathrm{vol} : \Delta\bm{E}\  \mathrm{d}V, \\     
		c(\delta p,\ \Delta p) & = 
		\int_{\Omega} \delta p \ \frac{1}{\kappa}\ \Delta p\  \mathrm{d}V.   
	\end{align}
\end{subequations}
It can be recognized that the volumetric part of the second Piola-Kirchhoff stress for this formulation is given as
\begin{equation}
	\bm{S}_\mathrm{vol} = p\, \bm{S}'_\mathrm{vol} = 2\, p\, \frac{\partial J}{\partial \bm{C}} = p\, J\, \bm{C}^{-1}
\end{equation} 
Note that this is in accordance with the derivation in \eqref{A1_Svol}, and the formulation can therefore be used with the proposed elasto-plastic material law. Note that due to the linearization of the problem, the volumetric and deviatoric parts of the stress and material tangent must not be determined. Within the implementation, the full material tensor and stress can therefore be used directly.

\subsection{Definition of the transformation matrix}
\label{t_matrix}
The transformation matrix $\bm{T}$ contains the entries of the inverse Jacobian $\bm{J}^{-1} = \bm{j}$ and is introduced in Section~\ref{Kinematics}. It is defined as follows in Nye's notation
\begin{equation}
    \renewcommand{\arraystretch}{1.0}
    \bm{T} =   \begin{pmatrix}
      j_{11}^2 & j_{21}^2 & j_{31}^2 & j_{11}\,j_{21} & j_{21}\,j_{31} & j_{11}\,j_{31} \\
      j_{12}^2 & j_{22}^2 & j_{32}^2 & j_{12}\,j_{22} & j_{22}\,j_{32} & j_{12}\,j_{32} \\
      j_{13}^2 & j_{23}^2 & j_{33}^2 & j_{13}\,j_{23} & j_{23}\,j_{33} & j_{13}\,j_{33} \\
      2\,j_{11}\,j_{12} & 2\,j_{21}\,j_{22} & 2\,j_{31}\,j_{32} & j_{12}\,j_{21} + j_{11}\,j_{22} & j_{22}\,j_{31} + j_{21}\,j_{32} & j_{12}\,j_{31} + j_{11}\,j_{32} \\
      2\,j_{11}\,j_{13} & 2\,j_{21}\,j_{23} & 2\,j_{31}\,j_{33} & j_{13}\,j_{21} + j_{11}\,j_{23} & j_{23}\,j_{31} + j_{21}\,j_{33} & j_{13}\,j_{31} + j_{11}\,j_{33}  \\
      2\,j_{12}\,j_{13} & 2\,j_{22}\,j_{23} & 2\,j_{32}\,j_{33} & j_{13}\,j_{22} + j_{12}\,j_{23} & j_{23}\,j_{32} + j_{22}\,j_{33} & j_{13}\,j_{32} + j_{12}\,j_{33} 
      \end{pmatrix}.
  \end{equation}
\subsection{Calculation of the residual vectors and stiffness matrices}
\label{res_stiff}
In accordance with Section~\ref{Gauss_point_concept}, both $\bm{B}_c$ and $\bm{B}_{\mathrm{enh}}$ obtain the following polynomial form
\begin{align}
  \begin{split}
  &\bm{B}_c = \bm{B}^{0}_c + \underbrace{\bm{B}^{\spacee\xi}_c\,\xi + \bm{B}^{\spacee\eta}_c\,\eta + \bm{B}^{\spacee\zeta}_c\,\zeta}_{=: \bm{B}^{\mathrm{hg1}}_c} + \underbrace{\bm{B}^{\spacee\xi\eta}_c\,\xi\eta + \bm{B}^{\spacee\xi\zeta}_c\,\xi\zeta + \bm{B}^{\spacee\eta\zeta}_c\,\eta\zeta}_{=: \bm{B}^{\mathrm{hg2}}_c}, \\
  &\bar{\bm{B}}_{\mathrm{enh}} = \bar{\bm{B}}_{\mathrm{enh}}^{\spacee\xi}\,\xi + \bar{\bm{B}}_{\mathrm{enh}}^{\spacee\eta}\,\eta + \bar{\bm{B}}_{\mathrm{enh}}^{\spacee\zeta}\,\zeta,
  \end{split}
\end{align}
while the superscript denotes the derivation with respect to the convective parameters. With this, the residual vectors and their corresponding stiffness matrices at element derived in Section~\ref{Discretization of the weak form} can be solved using analytical integration as follows 
\begin{align}
    \begin{split}
      \bm{R}_{w_e} &= \frac{8}{3} \,\det{(\bm{J}^0)}\Bigl((\bm{B}_{\mathrm{enh}}^{\spacee\xi})^T \bm{C}^{\mathrm{hg}}\hat{\bm{E}}^{\spacee\xi}_c + (\bm{B}_{\mathrm{enh}}^{\spacee\eta})^T\bm{C}^{\mathrm{hg}}\hat{\bm{E}}^{\spacee\eta}_c + (\bm{B}_{\mathrm{enh}}^{\spacee\zeta})^T\bm{C}^{\mathrm{hg}}\hat{\bm{E}}^{\spacee\zeta}_c \Bigr),\\
      \bm{R}^0_{u_e} &= 8 \, \det{(\bm{J}^0)}(\bm{B}^0_c)^T \hat{\bm{S}}^0 ,\\
      \bm{R}^{\mathrm{hg}}_{u_e} &= \frac{8}{3} \,\det{(\bm{J}^0)}\Bigl((\bm{B}^{\spacee\xi}_c)^T \bm{C}^{\mathrm{hg}}\hat{\bm{E}}^{\spacee\xi}_c + (\bm{B}^{\spacee\eta}_c)^T\bm{C}^{\mathrm{hg}}\hat{\bm{E}}^{\spacee\eta}_c + (\bm{B}^{\spacee\zeta}_c)^T\bm{C}^{\mathrm{hg}}\hat{\bm{E}}^{\spacee\zeta}_c \Bigr),\\
      &+ \frac{8}{9} \,\det{(\bm{J}^0)}\Bigl((\bm{B}^{\spacee\xi\eta}_c)^T \bm{C}^{\mathrm{hg}}\hat{\bm{E}}^{\spacee\xi\eta}_c + (\bm{B}^{\spacee\xi\zeta}_c)^T\bm{C}^{\mathrm{hg}}\hat{\bm{E}}^{\spacee\xi\zeta}_c + (\bm{B}^{\spacee\eta\zeta}_c)^T\bm{C}^{\mathrm{hg}}\hat{\bm{E}}^{\spacee\eta\zeta}_c\Bigr),\\
      \bm{K}_{ww_e} &= \frac{8}{3} \,\det{(\bm{J}^0)}\Bigl((\bm{B}_{\mathrm{enh}}^{\spacee\xi})^T\bm{C}^{\mathrm{hg}}\bm{B}_{\mathrm{enh}}^{\spacee\xi} + (\bm{B}_{\mathrm{enh}}^{\spacee\eta})^T\bm{C}^{\mathrm{hg}}\bm{B}_{\mathrm{enh}}^{\spacee\eta} + (\bm{B}_{\mathrm{enh}}^{\spacee\zeta})^T \bm{C}^{\mathrm{hg}}\bm{B}_{\mathrm{enh}}^{\spacee\zeta} \Bigr),\\
      \bm{K}_{uw_e} &= \frac{8}{3} \,\det{(\bm{J}^0)}\Bigl((\bm{B}^{\spacee\xi}_c)^T \bm{C}^{\mathrm{hg}}\bm{B}_{\mathrm{enh}}^{\spacee\xi} \,+\, (\bm{B}^{\spacee\eta}_c)^T\bm{C}^{\mathrm{hg}}\bm{B}_{\mathrm{enh}}^{\spacee\eta} + (\bm{B}^{\spacee\zeta}_c)^T\bm{C}^{\mathrm{hg}}\bm{B}_{\mathrm{enh}}^{\spacee\zeta} \Bigr),
    \end{split}
  \end{align}
while the relation
\begin{equation}
  \int_{V_e} (\cdot)\, \mathrm{d}V_e^0= \int_{-1}^{+1} \int_{-1}^{+1} \int_{-1}^{+1} (\cdot)\, \det{(\bm{J}^0)}\, \mathrm{d}\xi\,\mathrm{d}\eta\, \mathrm{d}\zeta\,
\end{equation}
holds.


\section{Appendix B. Declarations}
\label{appendix_B}

\subsection{Acknowledgements}
Hagen Holthusen, Sven Klinkel and Stefanie Reese gratefully acknowledge financial support of the project 495926269 within the research unit FOR 5492 by the Deutsche Forschungsgemeinschaft. Stefanie Reese gratefully acknowledges the financial support of the research work B05 within SFB/TRR 339 with the project number: 453596084. Hagen Holthusen gratefully acknowledges the financial support of the research work by the Deutsche Forschungsgemeinschaft (DFG, German Research Foundation) within the transregional Collaborative Research Center SFB/TRR 280, project-ID 417002380.

\subsection{Conflict of interest}
The authors of this work certify that they have no affiliations with or involvement in any organization or entity with any financial interest (such as honoraria; participation in speakers' bureaus; membership, employment, consultancies, stock ownership, or other equity interest; and expert testimony or patent-licensing arrangements), or non-financial interest (such as
personal or professional relationships, affiliations, knowledge or beliefs) in the subject matter or
materials discussed in this manuscript.

\subsection{Code availability}
Our source codes are accessible to the public at \url{https://doi.org/10.5281/zenodo.14259791}

\subsection{Contributions by the authors}
\textbf{Njomza Pacolli:} Conceptualization, Methodology, Software, Validation, Formal analysis, Investigation, Visualization, Writing – original draft, Writing – review and editing.
\textbf{Ahmad Awad:} Formal analysis, Investigation, Writing – original draft, Writing – review and editing.
\textbf{Jannick Kehls:} Writing – original draft, Writing – review and editing.
\textbf{Bjorn Sauren:}  Software, Formal analysis, Investigation, Writing – original draft, Writing – review and editing.
\textbf{Sven Klinkel:} Funding acquisition, Supervision, Writing – original draft, Writing – review and editing.
\textbf{Stefanie Reese:} Funding acquisition, Supervision, Writing – original draft, Writing – review and editing.
\textbf{Hagen Holthusen:} Conceptualization, Methodology, Software, Funding acquisition, Supervision, Writing – original draft, Writing – review and editing.

\subsection{Statement of AI-assisted tools usage}
This document was prepared with the assistance of OpenAI's ChatGPT, an AI language model. ChatGPT was used for language refinement. The authors reviewed, edited, and take full responsibility for the content and conclusions of this work.

\addtocontents{toc}{\vspace{2em}} 




\bibliographystyle{abbrvnat}

\begin{thebibliography}{80}
  \providecommand{\natexlab}[1]{#1}
  \providecommand{\url}[1]{\texttt{#1}}
  \expandafter\ifx\csname urlstyle\endcsname\relax
    \providecommand{\doi}[1]{doi: #1}\else
    \providecommand{\doi}{doi: \begingroup \urlstyle{rm}\Url}\fi
  
  \bibitem[Alves~de Sousa et~al.(2003)Alves~de Sousa, Natal~Jorge,
    Fontes~Valente, and C{\'e}sar~de S{\'a}]{sousa_new_2003}
  R.~Alves~de Sousa, R.~Natal~Jorge, R.~Fontes~Valente, and J.~C{\'e}sar~de
    S{\'a}.
  \newblock A new volumetric and shear locking-free 3d enhanced strain element.
  \newblock \emph{Engineering Computations}, 20\penalty0 (7):\penalty0 896--925,
    2003.
  \newblock \doi{10.1108/02644400310502036}.
  \newblock URL
    \url{https://www.emerald.com/insight/content/doi/10.1108/02644400310502036/full/html}.
  
  \bibitem[Ambati et~al.(2016)Ambati, Kruse, and
    De~Lorenzis]{ambati_phase-field_2016}
  M.~Ambati, R.~Kruse, and L.~De~Lorenzis.
  \newblock A phase-field model for ductile fracture at finite strains and its
    experimental verification.
  \newblock \emph{Computational Mechanics}, 57\penalty0 (1):\penalty0 149--167,
    2016.
  \newblock ISSN 0178-7675, 1432-0924.
  \newblock \doi{10.1007/s00466-015-1225-3}.
  \newblock URL \url{http://link.springer.com/10.1007/s00466-015-1225-3}.
  
  \bibitem[Areias et~al.(2003)Areias, C{\'e}sar~de S{\'a}, Ant{\'o}nio, and
    Fernandes]{areias_analysis_2003}
  P.~Areias, J.~C{\'e}sar~de S{\'a}, C.~C. Ant{\'o}nio, and A.~Fernandes.
  \newblock Analysis of 3d problems using a new enhanced strain hexahedral
    element.
  \newblock \emph{International Journal for Numerical Methods in Engineering},
    58\penalty0 (11):\penalty0 1637--1682, 2003.
  \newblock ISSN 1097-0207.
  \newblock \doi{10.1002/nme.835}.
  \newblock URL \url{https://onlinelibrary.wiley.com/doi/abs/10.1002/nme.835}.
  
  \bibitem[Armstrong et~al.(1966)Armstrong, Frederick,
    et~al.]{armstrong1966mathematical}
  P.~J. Armstrong, C.~O. Frederick, et~al.
  \newblock \emph{A mathematical representation of the multiaxial Bauschinger
    effect}, volume 731.
  \newblock Berkeley Nuclear Laboratories Berkeley, CA, 1966.
  
  \bibitem[Arunakirinathar and
    Reddy(1995{\natexlab{a}})]{arunakirinathar_further_1995}
  K.~Arunakirinathar and B.~Reddy.
  \newblock Further results for enhanced strain methods with isoparametric
    elements.
  \newblock \emph{Computer Methods in Applied Mechanics and Engineering},
    127\penalty0 (1):\penalty0 127--143, 1995{\natexlab{a}}.
  \newblock ISSN 0045-7825.
  \newblock \doi{10.1016/0045-7825(95)00845-0}.
  \newblock URL
    \url{https://www.sciencedirect.com/science/article/pii/0045782595008450}.
  
  \bibitem[Arunakirinathar and
    Reddy(1995{\natexlab{b}})]{arunakirinathar_geometrical_1995}
  K.~Arunakirinathar and B.~Reddy.
  \newblock Some geometrical results and estimates for quadrilateral finite
    elements.
  \newblock \emph{Computer methods in applied mechanics and engineering},
    122\penalty0 (3):\penalty0 307--314, 1995{\natexlab{b}}.
  \newblock ISSN 0045-7825.
  \newblock \doi{10.1016/0045-7825(94)00718-3}.
  \newblock URL
    \url{https://www.sciencedirect.com/science/article/pii/0045782594007183}.
  
  \bibitem[Barfusz et~al.(2021{\natexlab{a}})Barfusz, Brepols, van~der Velden,
    Frischkorn, and Reese]{barfusz_single_2021}
  O.~Barfusz, T.~Brepols, T.~van~der Velden, J.~Frischkorn, and S.~Reese.
  \newblock A single gauss point continuum finite element formulation for
    gradient-extended damage at large deformations.
  \newblock \emph{Computer Methods in Applied Mechanics and Engineering},
    373:\penalty0 113440, 2021{\natexlab{a}}.
  \newblock ISSN 00457825.
  \newblock \doi{10.1016/j.cma.2020.113440}.
  \newblock URL
    \url{https://linkinghub.elsevier.com/retrieve/pii/S0045782520306253}.
  
  \bibitem[Barfusz et~al.(2021{\natexlab{b}})Barfusz, van~der Velden, Brepols,
    Holthusen, and Reese]{barfusz_reduced_2021}
  O.~Barfusz, T.~van~der Velden, T.~Brepols, H.~Holthusen, and S.~Reese.
  \newblock A reduced integration-based solid-shell finite element formulation
    for gradient-extended damage.
  \newblock \emph{Computer Methods in Applied Mechanics and Engineering},
    382:\penalty0 113884, 2021{\natexlab{b}}.
  \newblock ISSN 00457825.
  \newblock \doi{10.1016/j.cma.2021.113884}.
  \newblock URL
    \url{https://linkinghub.elsevier.com/retrieve/pii/S0045782521002218}.
  
  \bibitem[Bartholomew-Biggs et~al.(2000)Bartholomew-Biggs, Brown, Christianson,
    and Dixon]{bartholomew-biggs_automatic_nodate}
  M.~Bartholomew-Biggs, S.~Brown, B.~Christianson, and L.~Dixon.
  \newblock Automatic differentiation of algorithms.
  \newblock \emph{Journal of Computational and Applied Mathematics}, 2000.
  \newblock URL
    \url{https://www.sciencedirect.com/science/article/pii/S0377042700004222}.
  
  \bibitem[Belytschko et~al.(1984)Belytschko, Ong, Liu, and
    Kennedy]{belytschko1984hourglass}
  T.~Belytschko, J.~S.-J. Ong, W.~K. Liu, and J.~M. Kennedy.
  \newblock Hourglass control in linear and nonlinear problems.
  \newblock \emph{Computer methods in applied mechanics and engineering},
    43\penalty0 (3):\penalty0 251--276, 1984.
  \newblock URL
    \url{https://www.sciencedirect.com/science/article/pii/0045782584900677}.
  
  \bibitem[Betsch and Stein(1999)]{betsch_numerical_1999}
  P.~Betsch and E.~Stein.
  \newblock Numerical implementation of multiplicative elasto-plasticity into
    assumed strain elements with application to shells at large strains.
  \newblock \emph{Computer methods in applied mechanics and engineering},
    179\penalty0 (3):\penalty0 215--245, 1999.
  \newblock ISSN 0045-7825.
  \newblock \doi{10.1016/S0045-7825(99)00063-8}.
  \newblock URL
    \url{https://www.sciencedirect.com/science/article/pii/S0045782599000638}.
  
  \bibitem[Betsch et~al.(1996)Betsch, Gruttmann, and Stein]{betsch_4-node_1996}
  P.~Betsch, F.~Gruttmann, and E.~Stein.
  \newblock A 4-node finite shell element for the implementation of general
    hyperelastic 3d-elasticity at finite strains.
  \newblock \emph{Computer Methods in Applied Mechanics and Engineering},
    130\penalty0 (1):\penalty0 57--79, 1996.
  \newblock ISSN 0045-7825.
  \newblock \doi{10.1016/0045-7825(95)00920-5}.
  \newblock URL
    \url{https://www.sciencedirect.com/science/article/pii/0045782595009205}.
  
  \bibitem[Bieber et~al.(2018)Bieber, Oesterle, Ramm, and
    Bischoff]{bieber_variational_2018}
  S.~Bieber, B.~Oesterle, E.~Ramm, and M.~Bischoff.
  \newblock A variational method to avoid locking—independent of the
    discretization scheme.
  \newblock \emph{International Journal for Numerical Methods in Engineering},
    114\penalty0 (8):\penalty0 801--827, 2018.
  \newblock ISSN 1097-0207.
  \newblock \doi{10.1002/nme.5766}.
  \newblock URL \url{https://onlinelibrary.wiley.com/doi/abs/10.1002/nme.5766}.
  
  \bibitem[Bieber et~al.(2023)Bieber, Auricchio, Reali, and
    Bischoff]{bieber_artificial_2023}
  S.~Bieber, F.~Auricchio, A.~Reali, and M.~Bischoff.
  \newblock Artificial instabilities of finite elements for nonlinear elasticity:
    Analysis and remedies.
  \newblock \emph{International Journal for Numerical Methods in Engineering},
    124\penalty0 (11):\penalty0 2638--2675, 2023.
  \newblock ISSN 1097-0207.
  \newblock \doi{10.1002/nme.7224}.
  \newblock URL \url{https://onlinelibrary.wiley.com/doi/abs/10.1002/nme.7224}.
  
  \bibitem[Bischoff and Ramm(1997)]{bischoff_shear_1997}
  M.~Bischoff and E.~Ramm.
  \newblock Shear deformable shell elements for large strains and rotations.
  \newblock \emph{International Journal for Numerical Methods in Engineering},
    40\penalty0 (23):\penalty0 4427--4449, 1997.
  \newblock ISSN 1097-0207.
  \newblock
    \doi{10.1002/(SICI)1097-0207(19971215)40:23<4427::AID-NME268>3.0.CO;2-9}.
  \newblock URL
    \url{https://onlinelibrary.wiley.com/doi/abs/10.1002/%28SICI%291097-0207%2819971215%2940%3A23%3C4427%3A%3AAID-NME268%3E3.0.CO%3B2-9}.
  
  \bibitem[B{\"o}hm et~al.(2023)B{\"o}hm, Korelc, Hudobivnik, Kraus, and
    Wriggers]{bohm_mixed_2023}
  C.~B{\"o}hm, J.~Korelc, B.~Hudobivnik, A.~Kraus, and P.~Wriggers.
  \newblock Mixed virtual element formulations for incompressible and
    inextensible problems.
  \newblock \emph{Computational Mechanics}, 72\penalty0 (6):\penalty0 1141--1174,
    2023.
  \newblock ISSN 1432-0924.
  \newblock \doi{10.1007/s00466-023-02340-9}.
  \newblock URL \url{https://doi.org/10.1007/s00466-023-02340-9}.
  
  \bibitem[Bonet and Bhargava(1995)]{bonet_uniform_1995}
  J.~Bonet and P.~Bhargava.
  \newblock A uniform deformation gradient hexahedron element with artificial
    hourglass control.
  \newblock \emph{International Journal for Numerical Methods in Engineering},
    38\penalty0 (16):\penalty0 2809--2828, 1995.
  \newblock ISSN 1097-0207.
  \newblock \doi{10.1002/nme.1620381608}.
  \newblock URL
    \url{https://onlinelibrary.wiley.com/doi/abs/10.1002/nme.1620381608}.
  
  \bibitem[Bradbury et~al.(2018)Bradbury, Frostig, Hawkins, Johnson, Leary,
    Maclaurin, Necula, Paszke, VanderPlas, Wanderman-Milne,
    et~al.]{bradbury2018jax}
  J.~Bradbury, R.~Frostig, P.~Hawkins, M.~J. Johnson, C.~Leary, D.~Maclaurin,
    G.~Necula, A.~Paszke, J.~VanderPlas, S.~Wanderman-Milne, et~al.
  \newblock Jax: composable transformations of python+ numpy programs.
  \newblock 2018.
  
  \bibitem[Büchter et~al.(1994)Büchter, Ramm, and
    Roehl]{buchter_three-dimensional_1994}
  N.~Büchter, E.~Ramm, and D.~Roehl.
  \newblock Three-dimensional extension of non-linear shell formulation based on
    the enhanced assumed strain concept.
  \newblock \emph{International journal for numerical methods in engineering},
    37\penalty0 (15):\penalty0 2551--2568, 1994.
  \newblock ISSN 1097-0207.
  \newblock \doi{10.1002/nme.1620371504}.
  \newblock URL
    \url{https://onlinelibrary.wiley.com/doi/abs/10.1002/nme.1620371504}.
  
  \bibitem[Cangiani et~al.(2015)Cangiani, Manzini, Russo, and
    Sukumar]{cangiani_hourglass_2015}
  A.~Cangiani, G.~Manzini, A.~Russo, and N.~Sukumar.
  \newblock Hourglass stabilization and the virtual element method.
  \newblock \emph{International Journal for Numerical Methods in Engineering},
    102\penalty0 (3):\penalty0 404--436, 2015.
  \newblock ISSN 1097-0207.
  \newblock \doi{10.1002/nme.4854}.
  \newblock URL \url{https://onlinelibrary.wiley.com/doi/abs/10.1002/nme.4854}.
  
  \bibitem[Cao et~al.(2002)Cao, Hu, Lu, Fukunaga, and Yao]{cao_3d_2002}
  Y.~P. Cao, N.~Hu, J.~Lu, H.~Fukunaga, and Z.~H. Yao.
  \newblock A 3d brick element based on hu–washizu variational principle for
    mesh distortion.
  \newblock \emph{International Journal for Numerical Methods in Engineering},
    53\penalty0 (11):\penalty0 2529--2548, 2002.
  \newblock ISSN 1097-0207.
  \newblock \doi{10.1002/nme.409}.
  \newblock URL \url{https://onlinelibrary.wiley.com/doi/abs/10.1002/nme.409}.
  
  \bibitem[Cardoso et~al.(2008)Cardoso, Yoon, Mahardika, Choudhry, Alves~de
    Sousa, and Fontes~Valente]{cardoso_enhanced_2008}
  R.~P.~R. Cardoso, J.~W. Yoon, M.~Mahardika, S.~Choudhry, R.~J. Alves~de Sousa,
    and R.~A. Fontes~Valente.
  \newblock Enhanced assumed strain ({EAS}) and assumed natural strain ({ANS})
    methods for one-point quadrature solid-shell elements.
  \newblock \emph{International Journal for Numerical Methods in Engineering},
    75\penalty0 (2):\penalty0 156--187, 2008.
  \newblock ISSN 1097-0207.
  \newblock \doi{10.1002/nme.2250}.
  \newblock URL \url{https://onlinelibrary.wiley.com/doi/abs/10.1002/nme.2250}.
  
  \bibitem[Cihan et~al.(2022)Cihan, Hudobivnik, Korelc, and
    Wriggers]{cihan_virtual_2022}
  M.~Cihan, B.~Hudobivnik, J.~Korelc, and P.~Wriggers.
  \newblock A virtual element method for 3d contact problems with non-conforming
    meshes.
  \newblock \emph{Computer Methods in Applied Mechanics and Engineering},
    402:\penalty0 115385, 2022.
  \newblock ISSN 0045-7825.
  \newblock \doi{10.1016/j.cma.2022.115385}.
  \newblock URL
    \url{https://www.sciencedirect.com/science/article/pii/S0045782522004534}.
  
  \bibitem[Coleman and Noll(1961)]{coleman_foundations_1961}
  B.~D. Coleman and W.~Noll.
  \newblock Foundations of linear viscoelasticity.
  \newblock \emph{Reviews of modern physics}, 33\penalty0 (2):\penalty0 239--249,
    1961.
  \newblock ISSN 0034-6861.
  \newblock \doi{10.1103/RevModPhys.33.239}.
  \newblock URL \url{https://link.aps.org/doi/10.1103/RevModPhys.33.239}.
  
  \bibitem[Crisfield et~al.(1995)Crisfield, Moita, Lyons, and
    Jeleni{\'c}]{crisfield_enhanced_1995}
  M.~Crisfield, G.~Moita, L.~Lyons, and G.~Jeleni{\'c}.
  \newblock Enhanced lower-order element formulations for large strains.
  \newblock \emph{Computational mechanics}, 17\penalty0 (1):\penalty0 62--73,
    1995.
  \newblock ISSN 1432-0924.
  \newblock \doi{10.1007/BF00356479}.
  \newblock URL \url{https://doi.org/10.1007/BF00356479}.
  
  \bibitem[Da~Veiga et~al.(2013)Da~Veiga, Brezzi, and Marini]{da2013virtual}
  L.~B. Da~Veiga, F.~Brezzi, and L.~D. Marini.
  \newblock Virtual elements for linear elasticity problems.
  \newblock \emph{SIAM Journal on Numerical Analysis}, 51\penalty0 (2):\penalty0
    794--812, 2013.
  \newblock URL \url{https://epubs.siam.org/doi/abs/10.1137/120874746}.
  
  \bibitem[de~Souza~Neto et~al.(1995)de~Souza~Neto, Peri{\'c}, Huang, and
    Owen]{de_souza_neto_remarks_1995}
  E.~de~Souza~Neto, D.~Peri{\'c}, G.~Huang, and D.~Owen.
  \newblock Remarks on the stability of enhanced strain elements in finite
    elasticity and elastoplasticity.
  \newblock \emph{Communications in Numerical Methods in Engineering},
    11\penalty0 (11):\penalty0 951--961, 1995.
  \newblock ISSN 1099-0887.
  \newblock \doi{10.1002/cnm.1640111109}.
  \newblock URL
    \url{https://onlinelibrary.wiley.com/doi/abs/10.1002/cnm.1640111109}.
  
  \bibitem[Dettmer and Reese(2004)]{dettmer_theoretical_2004}
  W.~Dettmer and S.~Reese.
  \newblock On the theoretical and numerical modelling of armstrong–frederick
    kinematic hardening in the finite strain regime.
  \newblock \emph{Computer Methods in Applied Mechanics and Engineering},
    193\penalty0 (1):\penalty0 87--116, 2004.
  \newblock ISSN 00457825.
  \newblock \doi{10.1016/j.cma.2003.09.005}.
  \newblock URL
    \url{https://linkinghub.elsevier.com/retrieve/pii/S0045782503005218}.
  
  \bibitem[Drucker and Prager(1952)]{drucker_soil_1952}
  D.~C. Drucker and W.~Prager.
  \newblock Soil mechanics and plastic analysis or limit design.
  \newblock \emph{Appl. Math}, 10\penalty0 (2):\penalty0 157--165, 1952.
  \newblock ISSN 0033-569X.
  \newblock URL \url{https://www.jstor.org/stable/43633942}.
  
  \bibitem[Frischkorn and Reese(2013)]{frischkorn_solid-beam_2013}
  J.~Frischkorn and S.~Reese.
  \newblock A solid-beam finite element and non-linear constitutive modelling.
  \newblock \emph{Computer Methods in Applied Mechanics and Engineering},
    265:\penalty0 195--212, 2013.
  \newblock ISSN 00457825.
  \newblock \doi{10.1016/j.cma.2013.06.009}.
  \newblock URL
    \url{https://linkinghub.elsevier.com/retrieve/pii/S0045782513001618}.
  
  \bibitem[Gain et~al.(2014)Gain, Talischi, and Paulino]{gain2014virtual}
  A.~L. Gain, C.~Talischi, and G.~H. Paulino.
  \newblock On the virtual element method for three-dimensional linear elasticity
    problems on arbitrary polyhedral meshes.
  \newblock \emph{Computer Methods in Applied Mechanics and Engineering},
    282:\penalty0 132--160, 2014.
  \newblock URL
    \url{https://www.sciencedirect.com/science/article/pii/S0045782514001509}.
  
  \bibitem[Glaser and Armero(1997)]{glaser_formulation_1997}
  S.~Glaser and F.~Armero.
  \newblock On the formulation of enhanced strain finite elements in finite
    deformations.
  \newblock \emph{Engineering Computations}, 14\penalty0 (7):\penalty0 759--791,
    1997.
  \newblock ISSN 0264-4401.
  \newblock \doi{10.1108/02644409710188664}.
  \newblock URL \url{https://doi.org/10.1108/02644409710188664}.
  
  \bibitem[Griewank and Walther(2008)]{griewank2008evaluating}
  A.~Griewank and A.~Walther.
  \newblock \emph{Evaluating derivatives: principles and techniques of
    algorithmic differentiation}.
  \newblock SIAM, 2008.
  
  \bibitem[Hao et~al.(2023)Hao, Yu, Wang, Yu, and Zhang]{hao_stabilized_2023}
  Q.~Hao, J.~Yu, X.~Wang, Y.~Yu, and B.~Zhang.
  \newblock Stabilized low-order finite-element formulation for static and
    dynamic simulation of saturated soils based on a hybrid integration scheme.
  \newblock \emph{Computers and Geotechnics}, 161:\penalty0 105596, 2023.
  \newblock ISSN 0266-352X.
  \newblock \doi{10.1016/j.compgeo.2023.105596}.
  \newblock URL
    \url{https://www.sciencedirect.com/science/article/pii/S0266352X23003531}.
  
  \bibitem[Hascoet and Pascual(2013)]{hascoet_tapenade_2013}
  L.~Hascoet and V.~Pascual.
  \newblock The tapenade automatic differentiation tool: Principles, model, and
    specification.
  \newblock \emph{ACM Transactions on Mathematical Software (TOMS)}, 39\penalty0
    (3):\penalty0 20:1--20:43, 2013.
  \newblock ISSN 0098-3500.
  \newblock \doi{10.1145/2450153.2450158}.
  \newblock URL \url{https://dl.acm.org/doi/10.1145/2450153.2450158}.
  
  \bibitem[Holthusen et~al.(2023)Holthusen, Rothkranz, Lamm, Brepols, and
    Reese]{holthusen_inelastic_2023}
  H.~Holthusen, C.~Rothkranz, L.~Lamm, T.~Brepols, and S.~Reese.
  \newblock Inelastic material formulations based on a co-rotated intermediate
    configuration—application to bioengineered tissues.
  \newblock \emph{Journal of the Mechanics and Physics of Solids}, 172:\penalty0
    105174, 2023.
  \newblock ISSN 00225096.
  \newblock \doi{10.1016/j.jmps.2022.105174}.
  \newblock URL
    \url{https://linkinghub.elsevier.com/retrieve/pii/S0022509622003507}.
  
  \bibitem[Hudobivnik and Korelc(2016)]{hudobivnik_closed-form_2016}
  B.~Hudobivnik and J.~Korelc.
  \newblock Closed-form representation of matrix functions in the formulation of
    nonlinear material models.
  \newblock \emph{Finite Elements in Analysis and Design}, 111:\penalty0 19--32,
    Apr. 2016.
  \newblock ISSN 0168-874X.
  \newblock \doi{10.1016/j.finel.2015.12.002}.
  \newblock URL
    \url{https://www.sciencedirect.com/science/article/pii/S0168874X15001845}.
  
  \bibitem[Hughes(1977)]{hughes_equivalence_1977}
  T.~J.~R. Hughes.
  \newblock Equivalence of finite elements for nearly incompressible elasticity.
  \newblock \emph{Journal of Applied Mechanics}, 44\penalty0 (1):\penalty0
    181--183, 1977.
  \newblock ISSN 0021-8936, 1528-9036.
  \newblock \doi{10.1115/1.3423994}.
  \newblock URL
    \url{https://asmedigitalcollection.asme.org/appliedmechanics/article/44/1/181/388644/Equivalence-of-Finite-Elements-for-Nearly}.
  
  \bibitem[Juhre and Reese(2010)]{juhre_reduced_2010}
  D.~Juhre and S.~Reese.
  \newblock A reduced integration finite element technology based on a
    thermomechanically consistent stabilisation for 3d problems.
  \newblock \emph{Computer methods in applied mechanics and engineering},
    199\penalty0 (29):\penalty0 2050--2058, 2010.
  \newblock ISSN 00457825.
  \newblock \doi{10.1016/j.cma.2010.03.004}.
  \newblock URL
    \url{https://linkinghub.elsevier.com/retrieve/pii/S0045782510000800}.
  
  \bibitem[Klinkel and Reichel(2019)]{klinkel_finite_2019}
  S.~Klinkel and R.~Reichel.
  \newblock A finite element formulation in boundary representation for the
    analysis of nonlinear problems in solid mechanics.
  \newblock \emph{Computer Methods in Applied Mechanics and Engineering},
    347:\penalty0 295--315, 2019.
  \newblock ISSN 0045-7825.
  \newblock \doi{10.1016/j.cma.2018.12.020}.
  \newblock URL
    \url{https://www.sciencedirect.com/science/article/pii/S0045782518306194}.
  
  \bibitem[Klinkel and Wagner(1997)]{klinkel_geometrical_1997}
  S.~Klinkel and W.~Wagner.
  \newblock A geometrical non-linear brick element based on the {EAS}-method.
  \newblock \emph{International Journal for Numerical Methods in Engineering},
    40\penalty0 (24):\penalty0 4529--4545, 1997.
  \newblock ISSN 1097-0207.
  \newblock
    \doi{10.1002/(SICI)1097-0207(19971230)40:24<4529::AID-NME271>3.0.CO;2-I}.
  \newblock URL
    \url{https://onlinelibrary.wiley.com/doi/abs/10.1002/%28SICI%291097-0207%2819971230%2940%3A24%3C4529%3A%3AAID-NME271%3E3.0.CO%3B2-I}.
  
  \bibitem[Klinkel et~al.(2006)Klinkel, Gruttmann, and
    Wagner]{klinkel_robust_2006}
  S.~Klinkel, F.~Gruttmann, and W.~Wagner.
  \newblock A robust non-linear solid shell element based on a mixed variational
    formulation.
  \newblock \emph{Computer methods in applied mechanics and engineering},
    195\penalty0 (1):\penalty0 179--201, 2006.
  \newblock ISSN 0045-7825.
  \newblock \doi{10.1016/j.cma.2005.01.013}.
  \newblock URL
    \url{https://www.sciencedirect.com/science/article/pii/S0045782505000435}.
  
  \bibitem[Korelc(1997)]{korelc_automatic_1997}
  J.~Korelc.
  \newblock Automatic generation of finite-element code by simultaneous
    optimization of expressions.
  \newblock \emph{Theoretical Computer Science}, 187\penalty0 (1):\penalty0
    231--248, Nov. 1997.
  \newblock ISSN 0304-3975.
  \newblock \doi{10.1016/S0304-3975(97)00067-4}.
  \newblock URL
    \url{https://www.sciencedirect.com/science/article/pii/S0304397597000674}.
  
  \bibitem[Korelc(2002)]{korelc_multi-language_nodate}
  J.~Korelc.
  \newblock Multi-language and multi-environment generation of nonlinear finite
    element codes.
  \newblock \emph{Engineering with computers}, 2002.
  \newblock URL \url{https://link.springer.com/article/10.1007/s003660200028}.
  
  \bibitem[Korelc and Wriggers(1996{\natexlab{a}})]{korelc_consistent_1996}
  J.~Korelc and P.~Wriggers.
  \newblock Consistent gradient formulation for a stable enhanced strain method
    for large deformations.
  \newblock \emph{Engineering Computations}, 13\penalty0 (1):\penalty0 103--123,
    1996{\natexlab{a}}.
  \newblock ISSN 0264-4401.
  \newblock \doi{10.1108/02644409610111001}.
  \newblock URL \url{https://doi.org/10.1108/02644409610111001}.
  
  \bibitem[Korelc and Wriggers(1996{\natexlab{b}})]{korelc_efficient_1996}
  J.~Korelc and P.~Wriggers.
  \newblock An efficient 3d enhanced strain element with taylor expansion of the
    shape functions.
  \newblock \emph{Computational Mechanics}, 19\penalty0 (2):\penalty0 30--40,
    1996{\natexlab{b}}.
  \newblock ISSN 1432-0924.
  \newblock \doi{10.1007/BF02757781}.
  \newblock URL \url{https://doi.org/10.1007/BF02757781}.
  
  \bibitem[Korelc and Wriggers(2016)]{korelc2016automation}
  J.~Korelc and P.~Wriggers.
  \newblock \emph{Automation of Finite Element Methods}.
  \newblock Springer, 2016.
  \newblock URL
    \url{https://link.springer.com/content/pdf/10.1007/978-3-319-39005-5.pdf}.
  
  \bibitem[Legay and Combescure(2003)]{legay_elastoplastic_2003}
  A.~Legay and A.~Combescure.
  \newblock Elastoplastic stability analysis of shells using the physically
    stabilized finite element shb8ps.
  \newblock \emph{International Journal for Numerical Methods in Engineering},
    57\penalty0 (9):\penalty0 1299--1322, 2003.
  \newblock ISSN 1097-0207.
  \newblock \doi{10.1002/nme.728}.
  \newblock URL \url{https://onlinelibrary.wiley.com/doi/abs/10.1002/nme.728}.
  
  \bibitem[Lion(2000)]{lion_constitutive_2000}
  A.~Lion.
  \newblock Constitutive modelling in finite thermoviscoplasticity: a physical
    approach based on nonlinear rheological models.
  \newblock \emph{International Journal of Plasticity}, 2000.
  \newblock URL
    \url{https://www.sciencedirect.com/science/article/pii/S0749641999000388}.
  
  \bibitem[Liu et~al.(1998)Liu, Guo, Tang, and
    Belytschko]{liu_multiple-quadrature_1998}
  W.~K. Liu, Y.~Guo, S.~Tang, and T.~Belytschko.
  \newblock A multiple-quadrature eight-node hexahedral finite element for large
    deformation elastoplastic analysis.
  \newblock \emph{Computer Methods in Applied Mechanics and Engineering},
    154\penalty0 (1):\penalty0 69--132, 1998.
  \newblock ISSN 00457825.
  \newblock \doi{10.1016/S0045-7825(97)00106-0}.
  \newblock URL
    \url{https://linkinghub.elsevier.com/retrieve/pii/S0045782597001060}.
  
  \bibitem[Macneal and Harder(1985)]{macneal_proposed_1985}
  R.~H. Macneal and R.~L. Harder.
  \newblock A proposed standard set of problems to test finite element accuracy.
  \newblock \emph{Finite elements in analysis and design}, 1\penalty0
    (1):\penalty0 3--20, 1985.
  \newblock ISSN 0168874X.
  \newblock \doi{10.1016/0168-874X(85)90003-4}.
  \newblock URL
    \url{https://linkinghub.elsevier.com/retrieve/pii/0168874X85900034}.
  
  \bibitem[Mohseni-Mofidi and Bierwisch(2021)]{mohseni-mofidi_application_2021}
  S.~Mohseni-Mofidi and C.~Bierwisch.
  \newblock Application of hourglass control to eulerian smoothed particle
    hydrodynamics.
  \newblock \emph{Computational Particle Mechanics}, 8\penalty0 (1):\penalty0
    51--67, 2021.
  \newblock ISSN 2196-4386.
  \newblock \doi{10.1007/s40571-019-00312-6}.
  \newblock URL \url{https://doi.org/10.1007/s40571-019-00312-6}.
  
  \bibitem[Pfefferkorn and Betsch(2023)]{pfefferkorn_hourglassing-_2023}
  R.~Pfefferkorn and P.~Betsch.
  \newblock Hourglassing- and locking-free mesh distortion insensitive
    petrov–galerkin {EAS} element for large deformation solid mechanics.
  \newblock \emph{International Journal for Numerical Methods in Engineering},
    124\penalty0 (6):\penalty0 1307--1343, 2023.
  \newblock ISSN 1097-0207.
  \newblock \doi{10.1002/nme.7166}.
  \newblock URL \url{https://onlinelibrary.wiley.com/doi/abs/10.1002/nme.7166}.
  
  \bibitem[Pfefferkorn et~al.(2021)Pfefferkorn, Bieber, Oesterle, Bischoff, and
    Betsch]{pfefferkorn_improving_2021}
  R.~Pfefferkorn, S.~Bieber, B.~Oesterle, M.~Bischoff, and P.~Betsch.
  \newblock Improving efficiency and robustness of enhanced assumed strain
    elements for nonlinear problems.
  \newblock \emph{International Journal for Numerical Methods in Engineering},
    122\penalty0 (8):\penalty0 1911--1939, 2021.
  \newblock ISSN 1097-0207.
  \newblock \doi{10.1002/nme.6605}.
  \newblock URL \url{https://onlinelibrary.wiley.com/doi/abs/10.1002/nme.6605}.
  
  \bibitem[Puso(2000)]{puso_highly_2000}
  M.~A. Puso.
  \newblock A highly efficient enhanced assumed strain physically stabilized
    hexahedral element.
  \newblock \emph{International Journal for Numerical Methods in Engineering},
    49\penalty0 (8):\penalty0 1029--1064, 2000.
  \newblock ISSN 1097-0207.
  \newblock \doi{10.1002/1097-0207(20001120)49:8<1029::AID-NME990>3.0.CO;2-3}.
  \newblock URL
    \url{https://onlinelibrary.wiley.com/doi/abs/10.1002/1097-0207(20001120)49:8%3C1029::AID-NME990%3E3.0.CO;2-3}.
  
  \bibitem[Reese(2003)]{reese_consistent_2003}
  S.~Reese.
  \newblock On a consistent hourglass stabilization technique to treat large
    inelastic deformations and thermo‐mechanical coupling in plane strain
    problems.
  \newblock \emph{International journal for numerical methods in engineering},
    57\penalty0 (8):\penalty0 1095--1127, 2003.
  \newblock ISSN 0029-5981, 1097-0207.
  \newblock \doi{10.1002/nme.719}.
  \newblock URL \url{https://onlinelibrary.wiley.com/doi/10.1002/nme.719}.
  
  \bibitem[Reese(2005)]{reese_physically_2005}
  S.~Reese.
  \newblock On a physically stabilized one point finite element formulation for
    three-dimensional finite elasto-plasticity.
  \newblock \emph{Computer Methods in Applied Mechanics and Engineering},
    194\penalty0 (45):\penalty0 4685--4715, 2005.
  \newblock ISSN 00457825.
  \newblock \doi{10.1016/j.cma.2004.12.012}.
  \newblock URL
    \url{https://linkinghub.elsevier.com/retrieve/pii/S0045782504005596}.
  
  \bibitem[Reese(2007)]{reese_large_2007}
  S.~Reese.
  \newblock A large deformation solid‐shell concept based on reduced
    integration with hourglass stabilization.
  \newblock \emph{International Journal for Numerical Methods in Engineering},
    69\penalty0 (8):\penalty0 1671--1716, 2007.
  \newblock ISSN 0029-5981, 1097-0207.
  \newblock \doi{10.1002/nme.1827}.
  \newblock URL \url{https://onlinelibrary.wiley.com/doi/10.1002/nme.1827}.
  
  \bibitem[Reese and Wriggers(2000)]{reese_stabilization_2000}
  S.~Reese and P.~Wriggers.
  \newblock A stabilization technique to avoid hourglassing in finite elasticity.
  \newblock \emph{International Journal for Numerical Methods in Engineering},
    48\penalty0 (1):\penalty0 79--109, 2000.
  \newblock ISSN 1097-0207.
  \newblock
    \doi{10.1002/(SICI)1097-0207(20000510)48:1<79::AID-NME869>3.0.CO;2-D}.
  \newblock URL
    \url{https://onlinelibrary.wiley.com/doi/abs/10.1002/%28SICI%291097-0207%2820000510%2948%3A1%3C79%3A%3AAID-NME869%3E3.0.CO%3B2-D}.
  
  \bibitem[Reese et~al.(1999)Reese, Küssner, and Reddy]{reese_new_1999}
  S.~Reese, M.~Küssner, and B.~D. Reddy.
  \newblock A new stabilization technique for finite elements in non-linear
    elasticity.
  \newblock \emph{International Journal for Numerical Methods in Engineering},
    44\penalty0 (11):\penalty0 1617--1652, 1999.
  \newblock ISSN 1097-0207.
  \newblock
    \doi{10.1002/(SICI)1097-0207(19990420)44:11<1617::AID-NME557>3.0.CO;2-X}.
  \newblock URL
    \url{https://onlinelibrary.wiley.com/doi/abs/10.1002/%28SICI%291097-0207%2819990420%2944%3A11%3C1617%3A%3AAID-NME557%3E3.0.CO%3B2-X}.
  
  \bibitem[Sauren and Klinkel(2024{\natexlab{a}})]{sauren_mesh_2024}
  B.~Sauren and S.~Klinkel.
  \newblock Mesh topology-based spurious pressure stabilization in 3d finite
    elasticity using voronoi tessellations.
  \newblock \emph{Computational Mechanics}, 2024{\natexlab{a}}.
  \newblock ISSN 1432-0924.
  \newblock \doi{10.1007/s00466-024-02558-1}.
  \newblock URL \url{https://doi.org/10.1007/s00466-024-02558-1}.
  
  \bibitem[Sauren and Klinkel(2024{\natexlab{b}})]{sauren_stability_2024}
  B.~Sauren and S.~Klinkel.
  \newblock On the stability of mixed polygonal finite element formulations in
    nonlinear analysis.
  \newblock \emph{International Journal for Numerical Methods in Engineering},
    125\penalty0 (9):\penalty0 e7358, 2024{\natexlab{b}}.
  \newblock ISSN 0029-5981, 1097-0207.
  \newblock \doi{10.1002/nme.7358}.
  \newblock URL \url{https://onlinelibrary.wiley.com/doi/10.1002/nme.7358}.
  
  \bibitem[Sauren et~al.(2023)Sauren, Klarmann, Kobbelt, and
    Klinkel]{sauren_mixed_2023}
  B.~Sauren, S.~Klarmann, L.~Kobbelt, and S.~Klinkel.
  \newblock A mixed polygonal finite element formulation for
    nearly-incompressible finite elasticity.
  \newblock \emph{Computer Methods in Applied Mechanics and Engineering},
    403:\penalty0 115656, 2023.
  \newblock ISSN 00457825.
  \newblock \doi{10.1016/j.cma.2022.115656}.
  \newblock URL
    \url{https://linkinghub.elsevier.com/retrieve/pii/S0045782522006119}.
  
  \bibitem[Schulz(1985)]{schulz_finite_1985}
  J.~C. Schulz.
  \newblock Finite element hourglassing control.
  \newblock \emph{International Journal for Numerical Methods in Engineering},
    21\penalty0 (6):\penalty0 1039--1048, 1985.
  \newblock ISSN 0029-5981, 1097-0207.
  \newblock \doi{10.1002/nme.1620210606}.
  \newblock URL \url{https://onlinelibrary.wiley.com/doi/10.1002/nme.1620210606}.
  
  \bibitem[Schwarze and Reese(2009)]{schwarze_reduced_2009}
  M.~Schwarze and S.~Reese.
  \newblock A reduced integration solid‐shell finite element based on the {EAS}
    and the {ANS} concept—geometrically linear problems.
  \newblock \emph{International Journal for Numerical Methods in Engineering},
    80\penalty0 (10):\penalty0 1322--1355, 2009.
  \newblock ISSN 0029-5981, 1097-0207.
  \newblock \doi{10.1002/nme.2653}.
  \newblock URL \url{https://onlinelibrary.wiley.com/doi/10.1002/nme.2653}.
  
  \bibitem[Schwarze and Reese(2011)]{schwarze_reduced_2011}
  M.~Schwarze and S.~Reese.
  \newblock A reduced integration solid‐shell finite element based on the {EAS}
    and the {ANS} concept—large deformation problems.
  \newblock \emph{International Journal for Numerical Methods in Engineering},
    85\penalty0 (3):\penalty0 289--329, 2011.
  \newblock ISSN 0029-5981, 1097-0207.
  \newblock \doi{10.1002/nme.2966}.
  \newblock URL \url{https://onlinelibrary.wiley.com/doi/10.1002/nme.2966}.
  
  \bibitem[Simo et~al.(1993)Simo, Armero, and Taylor]{simo_improved_1993}
  J.~Simo, F.~Armero, and R.~Taylor.
  \newblock Improved versions of assumed enhanced strain tri-linear elements for
    3d finite deformation problems.
  \newblock \emph{Computer methods in applied mechanics and engineering},
    110\penalty0 (3):\penalty0 359--386, 1993.
  \newblock ISSN 00457825.
  \newblock \doi{10.1016/0045-7825(93)90215-J}.
  \newblock URL
    \url{https://linkinghub.elsevier.com/retrieve/pii/004578259390215J}.
  
  \bibitem[Simo and Armero(1992)]{simo_geometrically_1992}
  J.~C. Simo and F.~Armero.
  \newblock Geometrically non‐linear enhanced strain mixed methods and the
    method of incompatible modes.
  \newblock \emph{International journal for numerical methods in engineering},
    33\penalty0 (7):\penalty0 1413--1449, 1992.
  \newblock ISSN 0029-5981, 1097-0207.
  \newblock \doi{10.1002/nme.1620330705}.
  \newblock URL \url{https://onlinelibrary.wiley.com/doi/10.1002/nme.1620330705}.
  
  \bibitem[Song(2018)]{song_scaled_2018}
  C.~Song.
  \newblock \emph{The Scaled Boundary Finite Element Method: Introduction to
    Theory and Implementation}.
  \newblock John Wiley \& Sons, 2018.
  \newblock ISBN 978-1-119-38815-9.
  
  \bibitem[Steinmann et~al.(2012)Steinmann, Hossain, and
    Possart]{steinmann_hyperelastic_2012}
  P.~Steinmann, M.~Hossain, and G.~Possart.
  \newblock Hyperelastic models for rubber-like materials: consistent tangent
    operators and suitability for treloar’s data.
  \newblock \emph{Archive of Applied Mechanics}, 82\penalty0 (9):\penalty0
    1183--1217, 2012.
  \newblock ISSN 0939-1533, 1432-0681.
  \newblock \doi{10.1007/s00419-012-0610-z}.
  \newblock URL \url{http://link.springer.com/10.1007/s00419-012-0610-z}.
  
  \bibitem[Taylor and Govindjee(2014)]{taylor_feap_nodate}
  R.~L. Taylor and S.~Govindjee.
  \newblock {FEAP}- a finite element analysis program.
  \newblock 2014.
  
  \bibitem[Valente et~al.(2004)Valente, Sousa, and Jorge]{valente_enhanced_2004}
  R.~A.~F. Valente, R.~J. A.~d. Sousa, and R.~M.~N. Jorge.
  \newblock An enhanced strain 3d element for large deformation elastoplastic
    thin-shell applications.
  \newblock \emph{Computational Mechanics}, 34\penalty0 (1):\penalty0 38--52,
    2004.
  \newblock ISSN 1432-0924.
  \newblock \doi{10.1007/s00466-004-0551-7}.
  \newblock URL \url{https://doi.org/10.1007/s00466-004-0551-7}.
  
  \bibitem[Veiga et~al.(2013)Veiga, Brezzi, Cangiani, Manzini, Marini, and
    Russo]{veiga_basic_2012}
  L.~Veiga, F.~Brezzi, A.~Cangiani, G.~Manzini, L.~Marini, and A.~Russo.
  \newblock Basic principles of virtual element methods.
  \newblock \emph{Mathematical Models and Methods in Applied Sciences}, 23, 2013.
  \newblock \doi{10.1142/S0218202512500492}.
  \newblock URL
    \url{https://www.worldscientific.com/doi/abs/10.1142/S0218202512500492}.
  
  \bibitem[Vladimirov and Reese(2008)]{vladimirov_anisotropic_2008}
  I.~N. Vladimirov and S.~Reese.
  \newblock Anisotropic finite plasticity with combined hardening and application
    to sheet metal forming.
  \newblock \emph{International Journal of Material Forming}, 1:\penalty0
    293--296, 2008.
  \newblock ISSN 1960-6206, 1960-6214.
  \newblock \doi{10.1007/s12289-008-0346-z}.
  \newblock URL \url{http://link.springer.com/10.1007/s12289-008-0346-z}.
  
  \bibitem[Voce(1955)]{voce1955practical}
  E.~Voce.
  \newblock A practical strain hardening function.
  \newblock \emph{Metallurgia}, 51:\penalty0 219--226, 1955.
  
  \bibitem[Wengert(1964)]{wengert1964simple}
  R.~E. Wengert.
  \newblock A simple automatic derivative evaluation program.
  \newblock \emph{Communications of the ACM}, 7\penalty0 (8):\penalty0 463--464,
    1964.
  \newblock URL \url{https://dl.acm.org/doi/10.1145/355586.364791}.
  
  \bibitem[Wriggers and Reese(1996)]{wriggers_note_1996}
  P.~Wriggers and S.~Reese.
  \newblock A note on enhanced strain methods for large deformations.
  \newblock \emph{Computer Methods in Applied Mechanics and Engineering},
    135\penalty0 (3):\penalty0 201--209, 1996.
  \newblock ISSN 0045-7825.
  \newblock \doi{10.1016/0045-7825(96)01037-7}.
  \newblock URL
    \url{https://www.sciencedirect.com/science/article/pii/0045782596010377}.
  
  \bibitem[Wriggers et~al.(2017)Wriggers, Reddy, Rust, and
    Hudobivnik]{wriggers_efficient_2017}
  P.~Wriggers, B.~D. Reddy, W.~Rust, and B.~Hudobivnik.
  \newblock Efficient virtual element formulations for compressible and
    incompressible finite deformations.
  \newblock \emph{Computational Mechanics}, 60\penalty0 (2):\penalty0 253--268,
    2017.
  \newblock ISSN 1432-0924.
  \newblock \doi{10.1007/s00466-017-1405-4}.
  \newblock URL \url{https://doi.org/10.1007/s00466-017-1405-4}.
  
  \bibitem[Wriggers et~al.(2020)Wriggers, Hudobivnik, and
    Aldakheel]{wriggers2020virtual}
  P.~Wriggers, B.~Hudobivnik, and F.~Aldakheel.
  \newblock A virtual element formulation for general element shapes.
  \newblock \emph{Computational Mechanics}, 66:\penalty0 963--977, 2020.
  \newblock URL
    \url{https://link.springer.com/article/10.1007/s00466-020-01891-5}.
  
  \bibitem[Zienkiewicz et~al.(2010)Zienkiewicz, Taylor, and
    Zhu]{zienkiewicz_finite_2010}
  O.~C. Zienkiewicz, R.~L. Taylor, and J.~Zhu.
  \newblock \emph{The finite element method: its basis and fundamentals}.
  \newblock Elsevier, Amsterdam Heidelberg, 6. ed., reprint., transferred to
    digital print edition, 2010.
  \newblock ISBN 978-0-7506-6320-5.
  
  \end{thebibliography}

\end{document}